\documentclass[numberedappendix]{emulateapj}

\slugcomment{Accepted for publication in AJ}

\shorttitle{Structural Parameters of seven SMC intermediate-age and old star clusters}
\shortauthors{Glatt et al.}

\begin{document}

\title{Structural Parameters of seven SMC intermediate-age and old star clusters
\altaffilmark{*}}

\author{Katharina Glatt\altaffilmark{1,2}, 
Eva K. Grebel\altaffilmark{1,2}, John S. Gallagher III.\altaffilmark{3},  
Antonella Nota\altaffilmark{4}, Elena Sabbi\altaffilmark{4}, 
Marco Sirianni\altaffilmark{4}, Gisella Clementini\altaffilmark{5}, 
Gary Da Costa\altaffilmark{6}, Monica Tosi\altaffilmark{5}, Daniel Harbeck\altaffilmark{3}, 
Andreas Koch\altaffilmark{7}, and Andrea Kayser\altaffilmark{1}}
\altaffiltext{*}{Based on observations made with the NASA/ESA Hubble Space Telescope, obtained 
at the Space Telescope Science Institute, which is operated by the Association of Universities 
for Research in Astronomy, Inc., under NASA contract NAS 5-26555. These observations are associated 
with program GO-10396.}
\altaffiltext{1}{Department of Physics, University of Basel, Klingelbergstrasse 82, 
CH-4056 Basel, Switzerland}
\altaffiltext{2}{Astronomisches Rechen-Institut, Zentrum f\"ur Astronomie der
Universit\"at Heidelberg, M\"onchhofstr.\ 12--14, D-69120 Heidelberg, Germany}
\altaffiltext{3}{Department of Astronomy, University of Wisconsin, 475 North 
Charter Street, Madison, WI 53706-1582}
\altaffiltext{4}{Space Telescope Science Institute, 3700 San Martin Drive, 
Baltimore, MD 21218}
\altaffiltext{5}{INAF - Osservatorio Astronomico di Bologna, Via Ranzani 1, 
40127 Bologna, Italy}
\altaffiltext{6}{Research School of Astronomy \& Astrophysics, The Australian National 
University, Mt Stromlo Observatory, via Cotter Rd, Weston, ACT 2611, Australia}
\altaffiltext{7}{Department of Physics and Astronomy, University of Leicester,  University 
Road, Leicester, LE1 7RH, UK }

\begin{abstract}
We present structural parameters for the seven intermediate-age and old star clusters 
NGC\,121, Lindsay\,1, Kron\,3, NGC\,339, NGC\,416, Lindsay\,38, and NGC\,419 in the Small 
Magellanic Cloud. We fit King profiles and Elson, Fall, and Freeman (EFF) profiles to both
surface-brightness and star count data taken with the Advanced Camera for Surveys aboard the 
Hubble Space Telescope. Clusters older than $\sim$1~Gyr show a spread in cluster core radii 
that increases with age, while the youngest clusters have relatively compact cores. No evidence 
for post core collapse clusters was found. We find no correlation between core radius and distance 
from the SMC center, although consistent with other studies of dwarf galaxies, some relatively old 
and massive clusters have low densities. The oldest SMC star cluster, the only globular NGC121, is 
the most elliptical object of the studied clusters. No correlation is seen between ellipticity 
and distance from the SMC center. The structures of these massive intermediate-age (1-8 Gyr) SMC 
star clusters thus appear to primarily result from internal evolutionary processes. 
\end{abstract}

\keywords{galaxies: star clusters, -- galaxies: Magellanic Clouds}

\section{Introduction}
The Small Magellanic Cloud (SMC) contains populous star clusters similar to those found in the Large 
Magellanic Cloud (LMC), although the two galaxies experienced a very different cluster formation 
history and age-metallicity relation \citep[e.g., ][]{daco02}. The smaller and less massive SMC contains 
many fewer clusters than the LMC. It formed its clusters continuously to the present day over the last 
$\sim$7.5~Gyr (age of Lindsay\,1, Glatt et al. 2008a). The oldest and only SMC globular star cluster, 
NGC\,121, is 2-3~Gyr younger than the oldest Milky Way (MW) and LMC globular clusters \citep{glatt08b}. 

Galactic globular clusters (GCs) can be described as an isothermal central region and a tidally 
truncated outer region \citep[e.g., ][]{binney98}, but both regions evolve with time. Once 
formed, star clusters are affected by internal and external processes, which influence the spatial 
distribution of member stars. The evolution of star clusters is affected by mass loss caused by, e.g.,
expulsion of gas, large-scale mass segregation, stellar mass loss, and low-mass star evaporation 
\citep[e.g., ][]{gnedin97,Koch04,lamers05,goodwin06}. The galactic environment causes external 
perturbations such as tidal shocking that occurs as star clusters cross the disk or pass near 
the bulge \citep[e.g., ][]{gnedin97}. These processes tend to decrease the cluster mass and 
might lead to core collapse, which has been observed in the oldest MW and LMC clusters 
\citep{djorgovski94,mackey03a}. The investigation of the present-day structure of individual star clusters is 
an important instrument to probe cluster dynamical evolution. 

In the SMC and the LMC, some of the older objects have apparently experienced a significant change in core radius,
while for other old objects the core radii apparently have remained almost unchanged \citep{mackey03a,mackey03b}. 
This former trend seems most likely to be the result of real physical cluster evolution, but 
the processes causing this core expansion for some clusters are not yet understood. A spread in
core radii beginning at a few 100~Myr is visible with a few clusters showing large core radii while
others remain small and compact. The five Fornax and four confirmed Sagittarius clusters show 
the same spread: two of the Sagittarius and two of the Fornax clusters have large core radii, while the 
others have compact cores \citep{mackey03c}. Galactic GCs show a spread in core radius size amounting
to about two orders of magnitude \citep{trager95} including a large number of so called core-collapse
clusters. However, many of the oldest GCs modified their original structure during their lifetime and have
developed small cores \citep[e.g., ][]{trager95,bonatto08}.

The ellipticity of the SMC clusters was noted to be larger than that of the MW and LMC clusters 
\citep{kontizas90,han94,goodwin97}. Old Galactic GCs have a very spherical shape, while the oldest LMC 
globulars are flatter. The oldest SMC clusters are even flatter than those in the LMC \citep{kontizas90}.

If one assumes that star clusters had originally small core regions and elliptical shapes then why was this 
original structure of many of the Magellanic Cloud (MS) GCs modified during their lifetime and why did some of the oldest 
clusters hosted by LMC and SMC remain unchanged? \citet{goodwin97} proposes that the strength of the tidal field of 
the parent galaxy is the dominant factor. If the tidal field is strong, the velocity anisotropies in a 
rapidly rotating elliptical globular cluster are destroyed, while a weak field is unable to remove these 
anisotropies and the cluster remains unchanged. In the MW, however, one has to distinguish between
halo-, bulge-, and disk GCs. Disk GCs move in circular orbits around the MW center and experience only little
variation of the Galactic gravitational field. Halo clusters pass the Galactic disk or bulge
\citep{hunter03,lamers05,gieles07}, which has a strong influence on their dymanical evolution and hence structure.
The GC system of the MW also contains a number of clusters acquired via merger processes \citep[e.g., ][]{bica06}.

We determine the structural parameters of the seven rich SMC star clusters NGC\,121, Lindsay\,1, 
Kron\,3, NGC\,339, NGC\,416, Lindsay\,38, and NGC\,419 by fitting {\it both} King and EFF profiles
to projected number-density and surface-brightness 
profiles. The observations were obtained with the Advanced Camera for Surveys (ACS) aboard the Hubble 
Space Telescope (HST). The important characteristic radii of star clusters that we determine
are the core radius ($r_c$), the projected half light radius ($r_h$), and the tidal radius ($r_t$). 
The core radius is defined as the radius at which the surface brightness has fallen to half its central value. 
The scale radius in the \citet{king62} analytic profile, $r_0$, however exceeds the core radius such that it 
is larger for lower central concentrations.
The difference between $r_c$ and $r_0$ is greatest at low concentrations. 
The half light radius contains half the light of the cluster. The mean SMC distance 
modulus is assumed to be $(m-M)_0$ = $18.88 \pm 0.1$~mag (60~kpc) \citep[e.g., ][]{storm04}, but our
photometry also provides us with individual cluster distances \citep[][Paper~I]{glatt08a}.

Generally, beyond the tidal radius the external gravitational fields of the galaxy dominate the internal dynamics, 
and stars are no longer bound to the cluster \citep[e.g., ][]{elson87,gnedin97}. 
In the MCs it is not obvious that the tidal field has set the observed tidal radii
of star clusters. For example \citet{elson87} found that ten LMC star clusters with ages up to 
$8 \times 10^8$ yrs do not appear to be tidally truncated. This could result, for example, from interactions
with other star clusters \citep{carvalho08}. Therefore, we use 'limiting radius' for the King model parameter 
$r_t$ because it may well be that the limiting radii are not tidally generated.   

In $\S$~\ref{sec:observation} we give an overview of the data and the reduction process, which has been 
described in detail in Paper~I and in \citet{glatt08b} (Paper~II). The methods used in the present 
paper are described in $\S$~\ref{sec:method} and the results are discussed in $\S$~\ref{sec:discussion}.

\begin{figure}
  \epsscale{1.2}
  \plotone{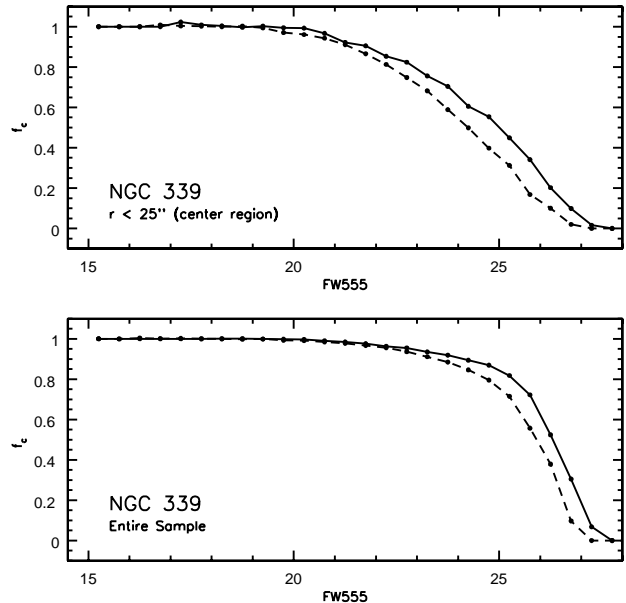}
  \caption{Completeness curves for the F555W (solid line) and F814W (dashed line) photometry of NGC\,339 as a 
  function of magnitude. The upper plot shows the completeness curves for the centre region (r $<$ 25'') of
  the cluster and the lower plot shows the completeness curves for the entire sample. Only stars for which the
  completeness in the center region is better than 50\% in the central regions of each cluster were used.}
\label{fig:cf}
\end{figure}

\section{Observations}
\label{sec:observation}

The SMC star clusters NGC\,121, Lindsay\,1, Kron\,3, NGC\,339, NGC\,416, Lindsay\,38, and NGC\,419
were observed with the HST/ACS between 2005 August and 2006 March (Table 1 in Paper I). The images
were taken in the F555W and F814W filters, which closely resemble the Johnson V and I filters
in their photometric properties \citep{siri05}. All clusters were observed with the Wide Field Channel 
(WFC), while for the dense central regions of NGC\,121, NGC\,416, and NGC\,419
images from the High Resolution Camera (HRC) are available. Each WFC image
covers an area of 200'' $\times$ 200'' at each pointing with a pixel scale of $\sim$0.05~arcsec. 
The HRC images cover an area of 29'' $\times$ 26'' with a pixel scale of $\sim$0.025~arcsec. 

The data sets were processed adopting the standard Space Telescope Science Institute ACS calibration 
pipeline (CALACS) to subtract the bias level and to apply the flat field correction. For each filter, 
the short and long exposures were co-added independently using the MULTIDRIZZLE package \citep{koek02}. 
Cosmic rays and hot pixels were removed with this package and a correction for geometrical distortion 
was provided. The resulting data consist of one 40~s and one 1984~s exposure (1940~s for Lindsay\,38) 
in F555W and one 20~s as well as one 1896~s exposure (1852~s for Lindsay\,38) in F814W. The HRC data of 
NGC\,419 consist of a 70~s and 1200~s exposure in F555W and a 40~s and 1036~s exposure in F814W.

Saturated foreground stars and background galaxies were discarded from the WFC images by using the 
Source Extractor \citep{bert96}. The detection thresholds were set at 3~$\sigma$ above the local 
background level for Lindsay\,1, 1~$\sigma$ for Kron\,3 and 4~$\sigma$ for NGC\,339, NGC\,416, and 
Lindsay\,38 in order to detect even the faintest sources. The threshold levels were chosen based on the 
different crowding effects of the individual clusters. The photometric reductions were carried out using 
the DAOPHOT package in the IRAF\footnote{\scriptsize{\small{IRAF} is distributed by the National Optical Astronomy 
Observatory, which is operated by the Association of Universities for Research in Astronomy, Inc. under cooperative 
agreement with the National Science Foundation.}}
environment on DRIZZLed images. The exposure times, the selection cuts, and the photometry are described 
in the Papers~I and~II and we refer to these two papers for detailed information.

\section{Structural Parameters}
\label{sec:method}
\subsection{Centers}

To study the structural parameters of the clusters we first determined the photo-center ($C_{phot}$) 
of the stellar populations; for a symmetric system this will be close to the center of gravity. 
An accurate determination of the cluster center is necessary in order 
to avoid artificial distortions of the radial profiles. As a first approximation we estimated the location of the
cluster center on the image by eye. A more precise center was then determined by calculating the average of 
the x and y coordinates within the cluster center region. First, the mean x and y coordinates were determined 
within a radius of 2000~pixels around the apparent center. In the following iterations, the radii were divided 
by two using the center coordinates found in the previous iterations as their origin. We iterated until 
the radius was smaller than 10 pixels. The error is $\sim$0.5'' for both $\alpha$ and $\delta$, which 
corresponds to 10 pixels in the HST/ACS images. For NGC\,121, NGC\,416, and NGC\,419 $C_{phot}$ was 
determined on the HRC data using the same algorithm due to crowding of the WFC data.
The resulting positions are summarized in Table~\ref{tab:center}. 

\begin{deluxetable*}{cccccc}
\tablecolumns{6}
\tablewidth{0pc}
\tablecaption{Position of the photo-center}
\tablenote{The values for the metallicities [Fe/H] are adopted from \citet{daco98} and Kayser et al. 2010,
in prep. The ages are taken from Paper~I and~II (best-fitting Dartmouth isochrones \citep{dotter07} 
for all clusters except NGC\,419, for which the Padova isochrones \citep{girardi00,girardi08} provided the best fit.)}
\tablehead{
\colhead{Cluster} & \colhead{$\alpha$} & \colhead{$\delta$} & \colhead{$[Fe/H]_{ZW84}$} & \colhead{age}  & \colhead{distance}\\
\colhead{} & \colhead{} & \colhead{} & \colhead{} & \colhead{Gyr} & \colhead{kpc} } 
\startdata
NGC\,121    & $0^h26^m47^s.80$ & $-71\arcdeg32'11''.40$ & $-1.46 \pm 0.10$ & $10.5 \pm 0.5$ & $64.9 \pm 1.2$ \\
Lindsay\,1  & $0^h03^m53^s.22$ & $-73\arcdeg28'16''.66$ & $-1.14 \pm 0.10$ & $7.5 \pm 0.5$   & $56.9 \pm 1.0$ \\
Kron\,3     & $0^h24^m46^s.28$ & $-72\arcdeg47'35''.76$ & $-1.08 \pm 0.12$ & $6.5 \pm 0.5$      & $60.6 \pm 1.1$ \\
NGC\,339    & $0^h57^m46^s.38$ & $-74\arcdeg28'14''.24$ & $-1.12 \pm 0.10$ & $6 \pm 0.5$     & $57.6 \pm 4.1$   \\
NGC\,416    & $1^h07^m58^s.64$ & $-72\arcdeg21'19''.75$ & $-1.00 \pm 0.13$ & $6 \pm 0.8$     & $60.4 \pm 1.9$  \\
Lindsay\,38 & $0^h48^m50^s.03$ & $-69\arcdeg52'07''.63$ & $-1.59 \pm 0.10$ & $6.5 \pm 0.5$  & $66.7 \pm 1.6$ \\
NGC\,419    & $1^h08^m17^s.31$ & $-72\arcdeg53'02''.49$ & $-0.67 \pm 0.12$ &  1.2-1.6	      & $50.2 \pm 2.6$  \\
\enddata
\label{tab:center}
\end{deluxetable*}

\subsection{King profile}

\begin{deluxetable*}{lcccccccc}
\tablecolumns{9}
\tablewidth{0pc}
\tablecaption{Structural Parameters from the King profile fit}
\tablehead{
\colhead{Cluster} & \colhead{$r_0$}  & \colhead{$r_0$\tablenotemark{1}} & \colhead{$r_t$}  & \colhead{$r_t$\tablenotemark{1}} & \colhead{c} & \colhead{$r_h$} & \colhead{$r_h$} & \colhead{$\phi$} \\
\colhead{}  & \colhead{arcsec} & \colhead{pc} & \colhead{arcsec} & \colhead{pc} & \colhead{} & \colhead{arcsec} & \colhead{pc} & \colhead{}} 
\startdata
\multicolumn{4}{l}{\it To number-density profiles} &&&&& \\
&&&&&&&& \\
NGC\,121     & $15.26 \pm 0.42$ & $4.80 \pm 0.56$   & $165.01 \pm 23.28$ & $51.92 \pm 7.32$  & $1.034 \pm 0.12$ & $27.01 \pm 2.21$ & $8.50 \pm 0.70$  & $10^{-3}$ \\
Lindsay\,1   & $61.67 \pm 3.80$ & $17.01 \pm 1.55$  & $230.77 \pm 37.26$ & $63.66 \pm 10.28$ & $0.573 \pm 0.10$ & $62.45 \pm 5.84$ & $17.23 \pm 1.61$ & $10^{-5}$ \\
Kron\,3      & $34.86 \pm 1.07$ & $10.24 \pm 0.87$  & $130.96 \pm 6.70$  & $38.47 \pm 1.97$  & $0.575 \pm 0.02$ & $35.38 \pm 1.50$ & $10.39 \pm 0.44$ & $10^{-2}$ \\
NGC\,339     & $32.84 \pm 0.64$ & $9.17 \pm 0.75$   & $186.73 \pm 16.51$ & $52.14 \pm 5.10$  & $0.755 \pm 0.06$ & $41.85 \pm 2.37$ & $11.69 \pm 0.66$ & $10^{-2}$ \\
NGC\,416     & $11.76 \pm 0.95$ & $3.44 \pm 0.44$   & $84.59 \pm 19.65$  & $24.77 \pm 5.75$  & $0.859 \pm 0.16$ & $16.96 \pm 2.63$ & $4.97 \pm 0.77$  & $10^{-2}$ \\
Lindsay\,38  & $31.24 \pm 0.85$ & $10.10 \pm 0.84$  & $173.82 \pm 12.39$ & $56.21 \pm 4.00$  & $0.745 \pm 0.04$ & $39.40 \pm 1.65$ & $12.74 \pm 0.53$ & $10^{-2}$ \\
NGC\,419     & $15.22 \pm 1.78$ & $3.70  \pm 0.51$  & $174.15 \pm 18.57$ & $42.38 \pm 4.52$  & $1.059 \pm 0.08$ & $27.69 \pm 2.47$ & $6.74 \pm 0.60$  & $10^{-2}$ \\
&&&&&&&& \\
\hline
&&&&&&&& \\
\multicolumn{4}{l}{\it To surface-brightness profiles} &&&&& \\
&&&&&&&& \\
NGC\,121   & $11.56 \pm 0.98$ & $3.64 \pm 0.31$   & $175.52 \pm 33.19$ & $55.22 \pm 10.44$ & $1.246 \pm 0.16$  & $24.15 \pm 3.20$ & $7.60 \pm 1.00$  & $10^{-5}$ \\
Lindsay\,1 & $61.41 \pm 4.95$ & $16.94 \pm 1.37$  & $216.54 \pm 41.92$ & $59.73 \pm 11.57$ & $0.349 \pm 0.08$  & $60.08 \pm 8.62$ & $16.57 \pm 2.37$ & $10^{-5}$  \\
Kron\,3    & $25.53 \pm 2.41$ & $7.50 \pm 0.71 $  & $180.25 \pm 37.56$ & $52.96 \pm 11.03$ & $0.848 \pm 0.15$  & $36.50 \pm 3.50$ & $10.72 \pm 1.03$ & $10^{-5}$  \\
NGC\,339   & $35.29 \pm 2.86$ & $9.86 \pm 0.80$   & $260.72 \pm 39.79$ & $72.80 \pm 11.11$ & $0.869 \pm 0.14$  & $51.58 \pm 5.89$ & $14.40 \pm 1.65$ & $10^{-5}$  \\
NGC\,416   & $10.22 \pm 1.51$ & $2.99 \pm 0.44$   & $107.95 \pm 22.54$ & $31.61 \pm  6.60$ & $1.023 \pm 0.15$  & $17.88 \pm 3.18$ & $5.24 \pm 0.93$  & $10^{-2}$  \\
Lindsay\,38& $31.47 \pm 3.68$ & $10.18 \pm 1.19$  & $179.65 \pm 34.08$ & $58.09 \pm 11.02$ & $0.711 \pm 0.05$  & $40.32 \pm 6.13$ & $13.04 \pm 1.98$ & $10^{-5}$  \\
NGC\,419   & $12.98 \pm 1.47$ & $3.16 \pm 0.36$   & $207.19 \pm 30.11$ & $50.42 \pm  7.33$ & $1.203 \pm 0.15$  & $27.73 \pm 3.58$ & $6.75 \pm 0.87$  & $10^{-5}$  \\
&&&&&&&& \\
\hline
&&&&&&&& \\
\multicolumn{6}{l}{\it To surface-brightness profiles for stars below the MSTOs} &&& \\
&&&&&&&& \\
NGC\,121    & $26.24 \pm 1.11$ & $8.26 \pm  0.35$  & $156.25 \pm 28.88$ & $49.16 \pm 9.09$  & $0.775 \pm 0.11$ & $34.31 \pm 3.91$ & $10.80 \pm 1.23$ & $10^{-2}$  \\
Lindsay\,1  & $78.21 \pm 4.48$ & $21.57 \pm 1.24$  & $269.84 \pm 50.11$ & $74.44 \pm 13.82$ & $0.538 \pm 0.07$ & $75.58 \pm 9.83$ & $20.85 \pm 2.71$ & $10^{-5}$  \\
Kron\,3     & $30.50 \pm 1.48$ & $8.96 \pm  0.44$  & $157.86 \pm 30.84$ & $46.38 \pm 9.06$  & $0.714 \pm 0.08$ & $37.01 \pm 4.60$ & $10.87 \pm 1.35$ & $10^{-2}$ \\
NGC\,339    & $32.54 \pm 2.05$ & $9.09 \pm  0.57$  & $271.17 \pm 52.07$ & $75.72 \pm 14.55$ & $0.921 \pm 0.12$ & $50.59 \pm 5.92$ & $14.13 \pm 1.65$ & $10^{-2}$ \\
NGC\,416    & $15.10 \pm 0.89$ & $4.42 \pm  0.26$  & $76.01  \pm 18.66$ & $22.26 \pm 5.46$  & $0.702 \pm 0.12$ & $18.05 \pm 2.81$ & $5.29 \pm 0.82$  & $10^{-5}$ \\
Lindsay\,38 & $29.01 \pm 1.84$ & $9.38 \pm  0.60$  & $192.65 \pm 38.72$ & $62.30 \pm 12.52$ & $0.822 \pm 0.06$ & $40.16 \pm 5.29$ & $12.99 \pm 1.71$ & $10^{-2}$ \\
NGC\,419    & $15.60 \pm 1.66$ & $3.80 \pm  0.40$  & $275.91 \pm 53.94$ & $67.15 \pm 13.13$ & $1.247 \pm 0.12$ & $35.03 \pm 5.18$ & $8.53 \pm 1.26$  & $10^{-2}$  \\
\enddata
\tablenotetext{1}{The MSTOs and the conversion of $r_0$ is based on the distances found in Paper~I. The upper section of this table corresponds to
the Figures~\ref{fig:king}, and~\ref{fig:king419} (solid lines), the middle section to Figures~\ref{fig:eff} and~\ref{fig:sb419} (dashed lines) for which only stars brighter than the 
magnitude for which the completeness is 50\% in the central region of each cluster were used. The lower section corresponds to the Figures~\ref{fig:effmsto} 
and~\ref{fig:sb419msto} (dashed lines) for which only stars brighter than the magnitude for which the completeness is 50\% in the central region of each cluster and 
stars fainter than the MSTOs were used. The half-light radii were computed by estimating $L_{tot}$ from the King profiles.}
\label{tab:sum_king}
\end{deluxetable*}

\begin{deluxetable}{ccccc}
\tablecolumns{5}
\tablewidth{0pc}
\tablecaption{Estimate of the Absolute Magnitudes}
\tablenote{The distance moduli and the reddenings $E_{V-I}$ are taken from Paper~I. }
\tablehead{
\colhead{Cluster} & \colhead{$(m-M)_0$}  & \colhead{$E_{V-I}$}  & \colhead{$M_V^{King}$} & \colhead{$M_V^{EFF}$} \\
\colhead{} & \colhead{mag} & \colhead{mag} & \colhead{mag} & \colhead{mag}} 
\startdata
NGC\,121    & $19.06 \pm 0.03$ & 0.024 & $-8.51 \pm 0.15$ & $-8.37 \pm 0.14$  \\
Lindsay\,1  & $18.78 \pm 0.04$ & 0.024 & $-7.39 \pm 0.09$ & $-7.38 \pm 0.09$  \\
Kron\,3     & $18.91 \pm 0.04$ & 0.024 & $-7.75 \pm 0.15$ & $-8.10 \pm 0.12$  \\
NGC\,339    & $18.80 \pm 0.08$ & 0.040 & $-7.42 \pm 0.14$ & $-7.76 \pm 0.11$  \\
NGC\,416    & $18.90 \pm 0.07$ & 0.104 & $-8.03 \pm 0.11$ & $-$  \\
Lindsay\,38 & $19.12 \pm 0.05$ & 0.016 & $-5.08 \pm 0.19$ & $-5.49 \pm 0.21$  \\
NGC\,419    & $18.50 \pm 0.12$ & 0.080 & $-8.85 \pm 0.18$ & $-$  \\
\enddata
\label{tab:absoluteMag}
\end{deluxetable}

The number surface density profiles of old GCs are usually described by the empirical King models 
\citep{king62}: \\ 

\begin{equation}
n(r) = k \cdot \{1/[1+(r/r_0)^2]^{1/2}-(1/[1+(r_t/r_0)^2]^{1/2})\}^2 + \phi , \\ 
\end{equation}

where n(r) is the number of stars per unit area, $r_0$ is the King radius, and $r_t$ is the limiting 
radius of the cluster. The parameter $\phi$ was added for the background contamination. No adjacent field 
was observed to measure the background. The field-of-view of ACS is too small for a reliable measurement of 
the background on the image. As a result we cannot necessarily distinguish an extended halo around a cluster 
from the true stellar background. Hence, we treat $\phi$ as a fitting parameter and assume that the background 
density $\phi$ is constant for each cluster. Although our background estimates are formally quite good, our 
values of $\phi$ can be considered as being greater than or equal to the actual background. Both for the number 
density and surface-brightness distributions King profiles were fitted, which are summarized in 
Table~\ref{tab:sum_king} (upper and middle sections).

Due to the field-of-view of the WFC the limiting radii of our clusters lie outside the ACS image. Therefore, 
the limiting radius $r_t$ and as a consequence also the half-light radius $r_h$ cannot be directly measured. 
We give an estimate of the projected half-light radius $r_h$ by calculating the total luminosity from the King 
profiles. The values $r_0$ and $r_t$ can be used to calculate the concentration parameter $c = log(r_t/r_0)$. 
From $r_h$ we can give an estimate of the absolute magnitude $M_V$ of the star clusters by multiplying the 
flux within $r_h$ by 2. Using the distance moduli and the extinctions from Paper~I, $M_V$ can be
calculated. The result is summarized in Table~\ref{tab:absoluteMag}. 

Concentric annuli containing the same number of stars were constructed around $C_{phot}$. The radii and the 
enclosed number of stars depend on the richness of the clusters and on photometric incompleteness caused by crowding. 
The completeness corrections on the WFC images were determined for each cluster separately and applied to the 
number density calculation. The completeness factors were determined using the subroutine {\it addstar} in DAOPHOT
to simulate 1,000,000 artificial stars (in steps of $\approx$2500 stars) in each long exposure frame. For a 
detailed description of the procedure we refer the reader to \citet{sabbi07}. Figure~\ref{fig:cf} shows the completeness 
factor of NGC\,339 in each filter, defined as the percentage of the artificial stars successfully recovered compared 
with the total number of stars added to the data. Only stars brighter than the magnitude for which the completeness 
is 50\% in the central regions of each cluster were used.

\begin{figure}
  \epsscale{1.2}
  \plotone{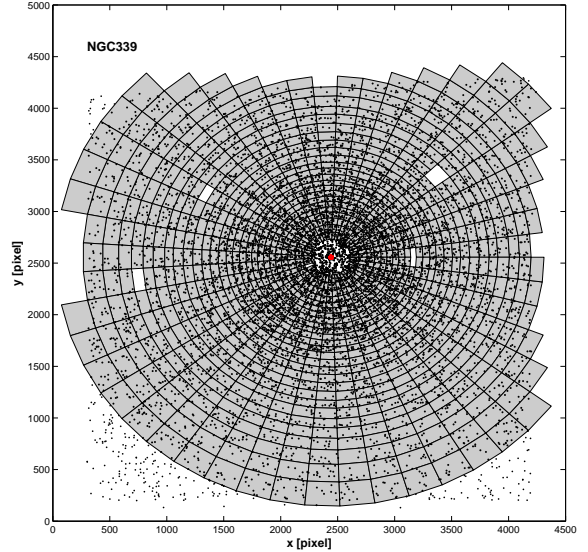}
  \caption{Star selection to calculate the number density profiles for NGC\,339 as this is representative
  for all clusters. Those stars lying within the gray area were counted while the grey areas represent the parts
  of the annuli in which these stars are found. The red dot indicates the location of the center of gravity.}
\label{fig:auswahl}
\end{figure}

Then the area of the annuli was calculated. We had to apply a geometrical area correction for those annuli that
were not fully imaged due to the cluster centering in the upper right part of the MULTIDRIZZLed image. 
Therefore, only those parts of the annuli were considered that lie fully on the image while the others
were excluded. Figure~\ref{fig:auswahl} displays the selected areas and stars of NGC\,339 as 
this is representative for all clusters. The black dots lying within the grey area represent the stars considered 
in the number density profile, while the filled grey areas show the annuli parts in which these stars were found.
The errors were propagated from the Poisson statistics of the number and area counts.
For NGC\,121, NGC\,416, and NGC\,419 the profiles were obtained on the WFC and HRC data independently and
then combined. The WFC data were only considered outside the HRC field, hence avoiding overlaps.
 
The star density was obtained by dividing the number of stars by their area. Both the King and the EFF profiles 
were fitted to each of our clusters via $\chi^2$ minimization. 

\begin{figure*}
  \epsscale{1.2}
  \plotone{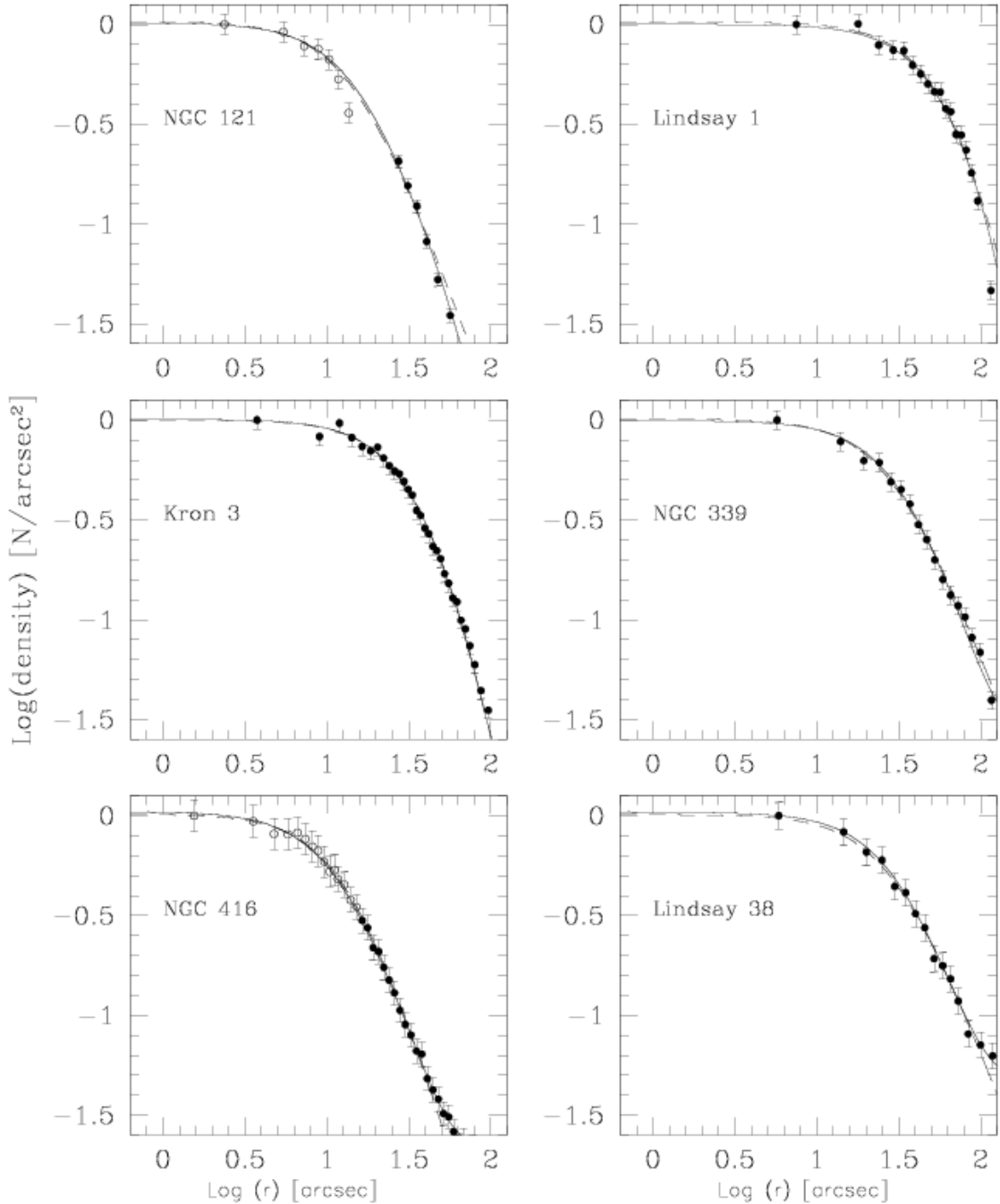}
  \caption{Number density profiles of the seven SMC clusters in our sample. The filled circles represent
  the WFC data, the open circles the HRC data. The solid line indicates the best-fitting King model and the
  dashed line the best-fitting EFF model of the radial 
  density distribution of the clusters. The radial plots of NGC\,416 and NGC\,419 (Fig.~\ref{fig:king419}) are 
  truncated.} 
\label{fig:king}
\end{figure*}

\begin{figure}
  \epsscale{1.1}
  \plotone{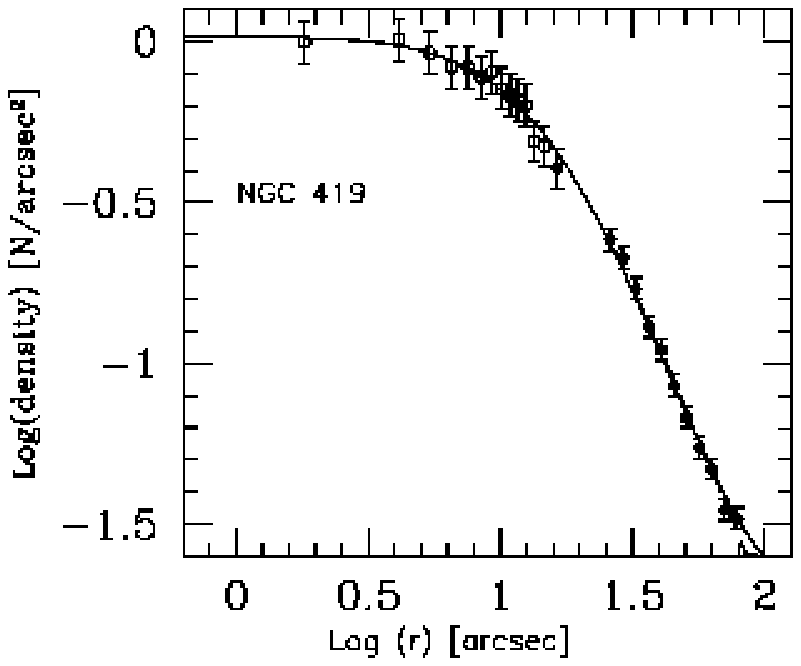}
  \caption{As for Fig.~\ref{fig:king} but for NGC\,419.}
\label{fig:king419}
\end{figure}

In Figure~\ref{fig:king} and~\ref{fig:king419} we show the stellar density distribution of our clusters with the 
best-fitting King profiles (solid line) and the best-fitting EFF profiles (dashed line) plus background.  
The filled circles represent the star density taken from the WFC data, the open circles the star density taken 
from the HRC data. The resulting structural parameters $r_0, r_t, r_h$, and c from the King profiles are 
summarized in Table~\ref{tab:sum_king}. The parameters $r_c, r_t$, $\gamma$, and $\phi$ from the EFF profiles 
are shown in Table~\ref{tab:sum_eff}. The listed errors are given by the $\chi^2$ minimization fitting process.

\subsection{EFF profile}

\begin{deluxetable*}{lcccccccc}
\tablecolumns{9}
\tablewidth{0pc}
\tablecaption{Structural Parameters from the EFF profile fit}
\tablehead{
\colhead{Cluster} & \colhead{$\mu_{555}(0)$} & \colhead{a} & \colhead{$r_c$} & \colhead{$r_c$\tablenotemark{1}} & \colhead{$\gamma$} & \colhead{$r_h$} & \colhead{$r_h$}  & \colhead{$\phi$} \\
\colhead{} & \colhead{$mag/arcsec^2$} & \colhead{arcsec} & \colhead{arcsec} & \colhead{pc} & \colhead{} & \colhead{arcsec} & \colhead{pc} & \colhead{} } 
\startdata
\multicolumn{4}{l}{\it To number-density profiles} &&&&& \\
&&&&&&&& \\
NGC\,121    & $18.87 \pm 0.03$ & $14.63 \pm 2.85$  & $13.42 \pm 2.62$  & $4.22 \pm 00.82$ & $2.27$ & $ - $             &  $ - $            & $10^{-5}$ \\
Lindsay\,1  & $21.25 \pm 0.04$ & $74.84 \pm 32.40$ & $48.53 \pm 19.00$ & $13.39 \pm 5.24$ & $3.95$ & $76.18 \pm 32.98$ &  $21.01 \pm 9.10$ & $10^{-5}$ \\
Kron\,3     & $20.56 \pm 0.03$ & $45.31 \pm 7.89$  & $28.66 \pm 4.99$  & $8.42 \pm 1.47$  & $4.12$ & $43.54 \pm 7.58$  &  $12.79 \pm 2.23$ & $10^{-5}$ \\
NGC\,339    & $20.03 \pm 0.03$ & $29.07 \pm 5.49$  & $27.58 \pm 5.21$  & $7.70 \pm 1.46$  & $2.16$ & $ - $             &  $ - $            & $10^{-5}$ \\
NGC\,416    & $18.21 \pm 0.06$ & $13.73 \pm 2.92$  & $11.27 \pm 2.40$  & $3.30 \pm 0.70$  & $2.69$ & $34.98 \pm 7.42$  &  $10.22 \pm 2.17$ & $10^{-2}$ \\
Lindsay\,38 & $22.90 \pm 0.04$ & $29.07 \pm 9.51$  & $27.58 \pm 5.02$  & $8.92 \pm 1.62$  & $2.16$ & $ - $             &  $ - $            & $10^{-5}$ \\
NGC\,419    & $18.18 \pm 0.03$ & $14.68 \pm 2.67$  & $14.20 \pm 2.52$  & $3.46 \pm 0.61$  & $1.05$ & $ - $             &  $ - $            & $10^{-5}$ \\
&&&&&&&& \\
\hline
&&&&&&&& \\
\multicolumn{4}{l}{\it To surface-brightness profiles} &&&&& \\
&&&&&&&& \\
NGC\,121    & $18.60 \pm 0.08$ & $16.52 \pm 0.76$  & $11.91 \pm 0.57$ & $3.75 \pm 0.39$  & $3.36$ & $22.66 \pm 0.99$ & $7.13  \pm 0.31$ & $10^{-5}$ \\
Lindsay\,1  & $21.94 \pm 0.09$ & $101.15 \pm 4.52$ & $50.05 \pm 1.96$ & $13.81 \pm 0.60$ & $6.33$ & $62.14 \pm 1.76$ & $17.14 \pm 0.48$ & $10^{-5}$ \\
Kron\,3     & $20.13 \pm 0.09$ & $30.11 \pm 1.35$  & $22.79 \pm 0.78$ & $6.70 \pm 0.26$  & $3.06$ & $49.47 \pm 1.41$ & $14.53 \pm 0.33$ & $10^{-5}$ \\
NGC\,339    & $21.31 \pm 0.10$ & $38.71 \pm 1.52$  & $29.02 \pm 1.19$ & $8.10 \pm 0.11$  & $3.11$ & $61.06 \pm 2.10$ & $17.05 \pm 0.45$ & $10^{-5}$ \\
NGC\,416    & $18.31 \pm 0.04$ & $9.20 \pm 0.50$   & $9.04 \pm 0.35$  & $2.65 \pm 0.07$  & $2.05$ & - & -                               & $1.8 \times 10^{-2}$ \\
Lindsay\,38 & $23.21 \pm 0.06$ & $37.83 \pm 1.47$  & $28.76 \pm 1.17$ & $9.30 \pm 0.48$  & $3.04$ & $64.71 \pm 1.84$ & $20.93 \pm 0.55$ & $10^{-5}$ \\
NGC\,419    & $17.70 \pm 0.06$ & $12.33 \pm 0.56$  & $11.73 \pm 0.47$ & $2.85 \pm 0.09$  & $2.15$ & - & -                               & $10^{-2}$ \\
&&&&&&&& \\
\hline
&&&&&&&& \\
\multicolumn{6}{l}{\it To surface-brightness profiles for stars below the MSTOs} &&& \\
&&&&&&&& \\
NGC\,121    & $21.90 \pm 0.07$ & $35.59 \pm 0.86$  & $23.46 \pm 0.59$ & $7.38 \pm 0.21$  & $3.99$ & $35.72 \pm 0.86$ & $11.24 \pm 0.27$ & $10^{-5}$ \\
Lindsay\,1  & $21.89 \pm 0.10$ & $111.84 \pm 3.61$ & $64.48 \pm 2.29$ & $17.79 \pm 0.63$ & $5.11$ & $83.83 \pm 2.70$ & $23.13 \pm 0.74$ & $10^{-5}$ \\
Kron\,3     & $21.79 \pm 0.08$ & $34.17 \pm 0.17$  & $25.31 \pm 0.48$ & $7.44 \pm 0.12$  & $3.20$ & $50.40 \pm 0.95$ & $14.81 \pm 0.28$ & $10^{-5}$ \\
NGC\,339    & $22.37 \pm 0.11$ & $29.90 \pm 0.18$  & $28.32 \pm 1.35$ & $7.91 \pm 0.38$  & $2.19$ & $ - $            & $ - $            & $10^{-5}$ \\
NGC\,416    & $20.91 \pm 0.06$ & $18.91 \pm 0.56$  & $17.08 \pm 0.51$ & $5.00 \pm 0.16$  & $2.42$ & $96.70 \pm 2.86$ & $28.32 \pm 0.83$ & $1.8 \times 10^{-2}$ \\
Lindsay\,38 & $24.67 \pm 0.09$ & $25.80 \pm 1.88$  & $25.20 \pm 1.71$ & $8.15 \pm 0.59$  & $2.07$ & $ - $            & $ - $            & $10^{-5}$ \\
NGC\,419    & $11.62 \pm 0.56$  & $14.37 \pm 0.49$ & $17.65 \pm 0.58$ & $3.50 \pm 0.10$  & $1.56$ & $ - $            & $ - $            & $10^{-2}$ \\
\enddata
\tablenotetext{1}{The conversion of $r_c$ is based on the distances found in Paper~I. The upper section of this table corresponds to the Figures~\ref{fig:eff} 
and~\ref{fig:sb419} (solid lines), the middle section to the Figures~\ref{fig:king} and~\ref{fig:king419} (dashed lines) for which only stars brighter than 
the magnitude for which the completeness is 50\% in the central region of each cluster were used. The third section of this table corresponds to the 
Figures~\ref{fig:effmsto} and~\ref{fig:sb419msto} (solid lines) for which only stars brighter than the magnitude for which the completeness is 50\% in the 
central region of each cluster and stars fainter than the MSTOs were used. The half-light radii were computed by estimating $L_{tot}$ from the EFF profiles. 
For the clusters with a $\gamma$ close to two a formally infinite model luminosity was derived and no half-light radius could be calculated.}
\label{tab:sum_eff}
\end{deluxetable*}

The EFF profile \citep{elson87} is given by \\ 

\begin{equation}
n(r) = n_0 \cdot (1 + r^2/a^2)^{-\gamma/2} + \phi, \\ 
\end{equation}

where $n_0$ is the central projected stellar density, $a$ is a measure of the core radius and $\gamma$ 
is the power-law slope. The parameter $\phi$ was added for the background contamination. The parameter 
$a$ in the EFF profiles is related to the core radius of a cluster by $r_c = a \cdot (2^{2/\gamma} 
- 1)^{1/2}$. To determine the parameters of the clusters we fitted the surface brightness profiles with 
$I(r) =I_0 \cdot [(1 + r^2/a^2)^{-\gamma/2}] + \phi$. No limiting radii can be derived from EFF profiles. We 
give an estimate of the projected half-light radius $r_h$ by calculating the total luminosity from the EFF profiles. This 
method works well until $\gamma \approx$ 2 is reached, because then the total luminosity formally becomes infinite. 

The EFF profiles were used in this study, because all recent studies \citep{kontizas82,kontizas83, 
kontizas86,kontizas90,mackey03b,hill06,carvalho08} of structure parameters of SMC star clusters used 
EFF profiles. This choice facilitates the comparison of our results with the earlier studies. For the 
same reason we chose to work in the commonly used F555W ($\sim$V) band. 
 
In order to measure the surface-brightness profiles, concentric annuli of coextensive areas were 
created around $C_{phot}$. The surface brightness $\mu_i$ of the \textit{i}th annulus in a set was found by 
summing over the flux of all stars that fall into the annulus. For NGC\,416, the profile was fitted only using the 
inner points (log(r)$<$1.5~arcsec), because the SMC field contribution is not negligible for this cluster. 
The same area and completeness corrections as for the number 
density profiles were applied. Saturated foreground stars were removed from the images. Only stars brighter 
than the magnitude for which the completeness is 50\% in the central region of each cluster were used.

\begin{figure*}
  \epsscale{1.2}
  \plotone{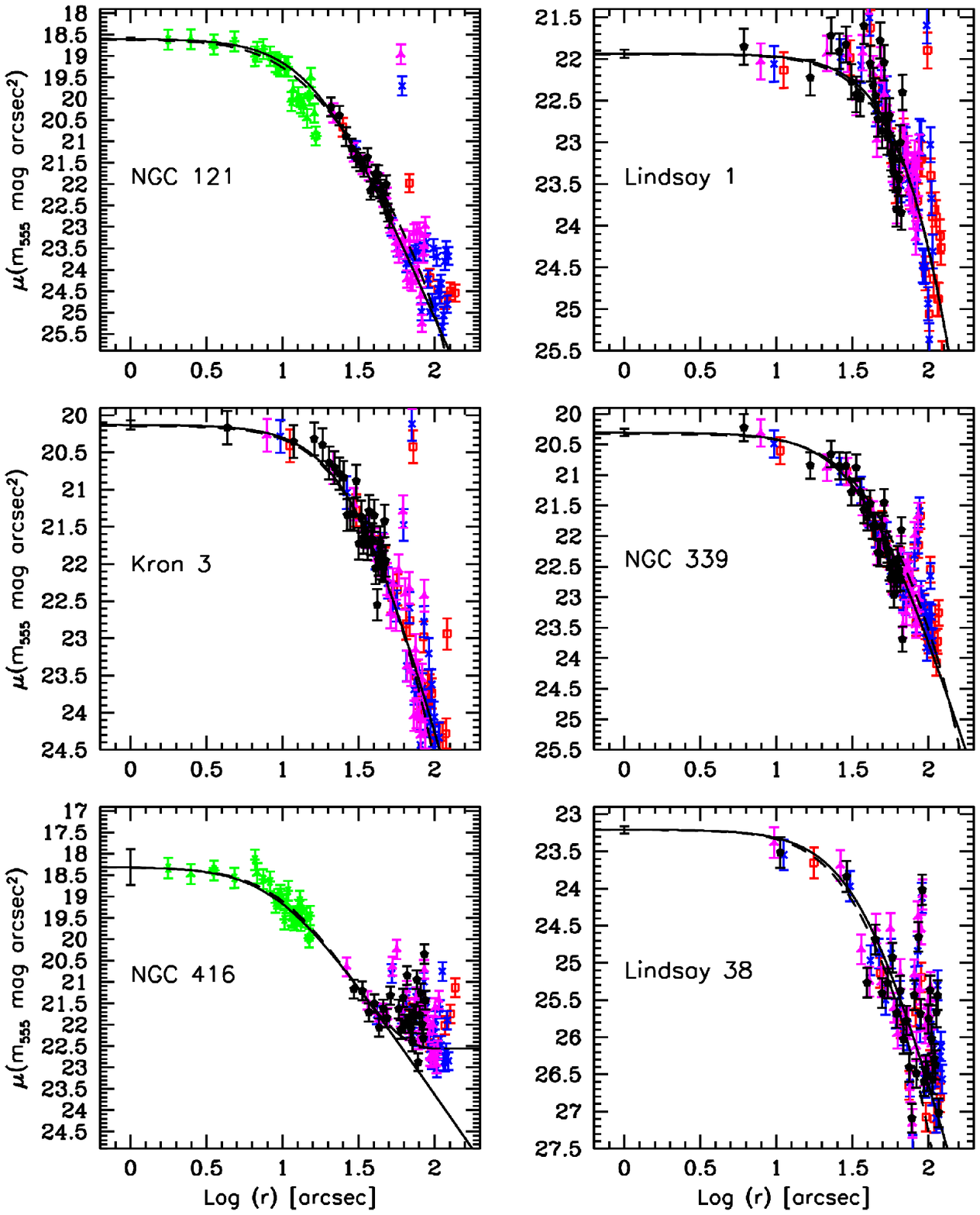}
  \caption{Surface-brightness profiles for each of the seven clusters in the present sample in the F555W-band. Only cluster 
  stars brighter than the magnitude for which the completeness is 50\% in the central region of each cluster were used.
  The surface-brightness was measured in four different areas which we display using different symbols (and different
  colors). For Lindsay\,1, Kron\,3, NGC\,339, and Lindsay\,38 only WFC images were used, while for NGC\,121, 
  NGC\,416, and NGC\,419 (Fig.~\ref{fig:sb419}) the dense center regions are covered with HRC data. The solid line
  indicates the best-fitting EFF model and the dashed line the best-fitting King model of the surface-brightness distribution.}
\label{fig:eff}
\end{figure*}

For Lindsay\,1, Kron\,3, NGC\,339, and Lindsay\,38 only WFC photometry is available, which resolves the clusters 
entirely. The surface brightness measured in four sets of concentric annuli with the same area are plotted for 
each cluster in Figure~\ref{fig:eff}. The first set was done with 76~arcsec$^2$ (pentagons, black), 
the second with 250~arcsec$^2$ (triangles, magenta), the third with 374~arcsec$^2$ (crosses, blue), and the fourth 
with 500~arcsec$^2$ (squares, red). 

For the three densest clusters, NGC\,121, NGC\,416, and NGC\,419, the central regions were observed 
with HST/HRC. The surface brightness profiles of the center region were determined using the HRC data, while
the profiles of the outer regions were calculated using the WFC data. Three HRC sets measured in concentric annuli 
of the same area are displayed with different symbols (green): 37.5~$arcsec^2$ (triangles), 25~$arcsec^2$ (asterisks), 
and 50~$arcsec^2$ (circles). For the WFC data, the same areas as above have been used, but only those annuli are 
shown that lie outside the regions covered by the HRC data. The surface-brightness profiles were fitted via $\chi^2$ 
minimization. 

\begin{figure}
  \epsscale{1.1}
  \plotone{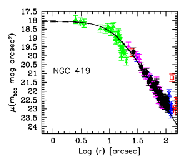}
  \caption{As for Fig.~\ref{fig:eff} but for NGC\,419.}
\label{fig:sb419}
\end{figure}

Two sets of EFF profiles were fitted. For the first set, all stars brighter than the magnitude for which the 
completeness is 50\% in the central region of each cluster were used. The resulting parameters can be compared with 
literature values. The best-fitting EFF profiles are shown in Figures~\ref{fig:eff} and~\ref{fig:sb419} (solid lines). 
The results are summarized in Table~\ref{tab:sum_eff} (upper and middle section).

\begin{figure*}
  \epsscale{1.2}
  \plotone{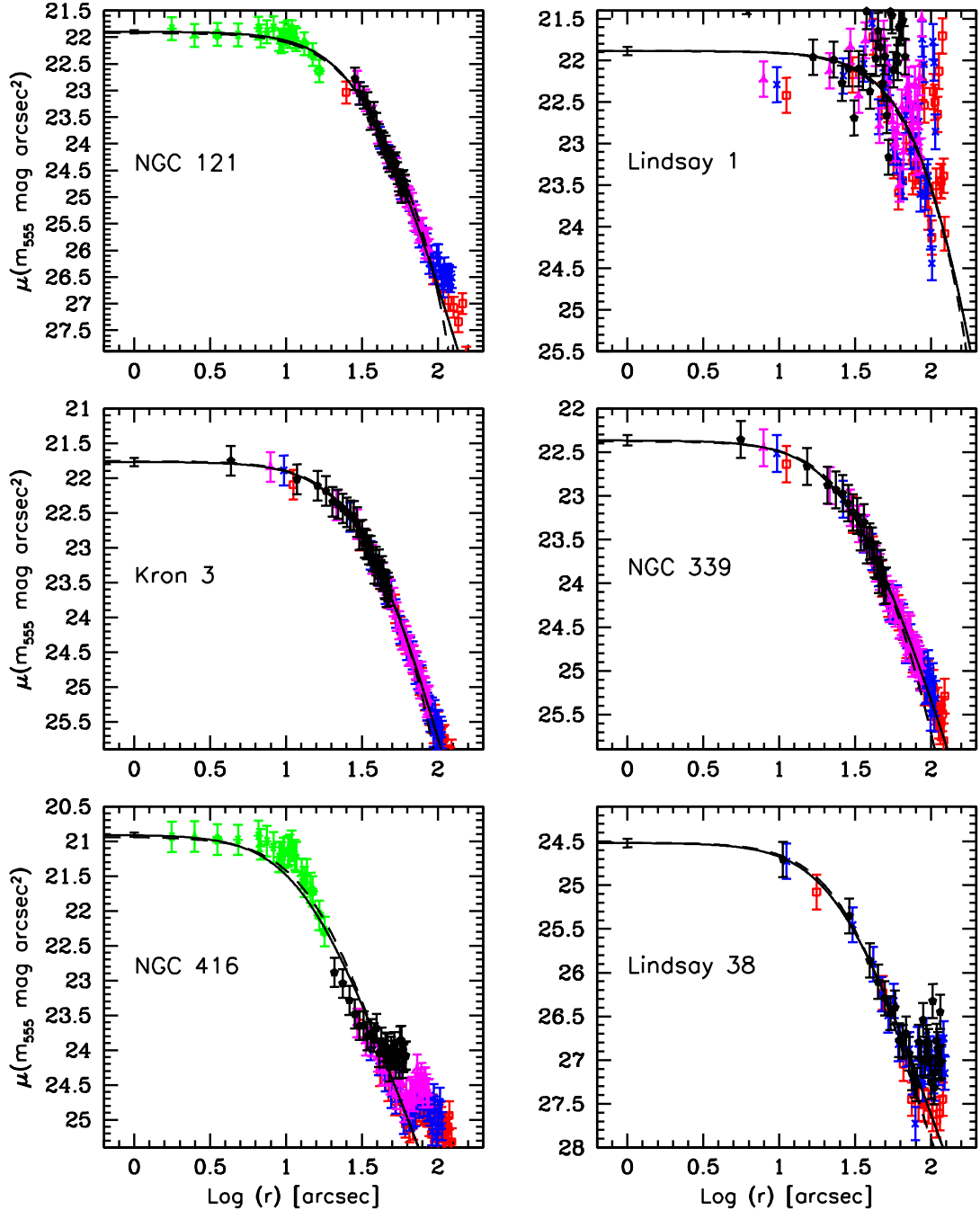}
  \caption{For this set of profiles all cluster stars brighter than the magnitude for which the completeness is 50\% in the
  cluster central region and fainter than the MSTO were used. The applied method is the same as for the first set of EFF 
  profiles shown in Fig.~\ref{fig:eff} and~\ref{fig:sb419}.}
\label{fig:effmsto}
\end{figure*}

\begin{figure}
  \epsscale{1.1}
  \plotone{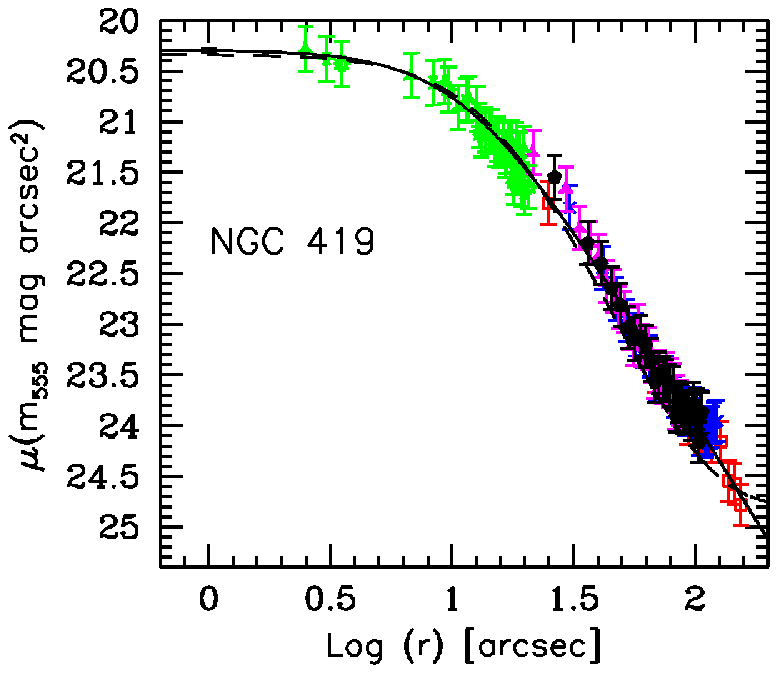}
  \caption{As for Fig.~\ref{fig:effmsto} but for NGC\,419.}
\label{fig:sb419msto}
\end{figure}

The surface brightness comes mostly from the brighter stars around the main-sequence turnoff (MSTO) and brighter, 
while the surface density distribution comes principally from the numerous stars below the MSTO. Due to the long 2-body 
relaxation times, we do not expect much mass segregation and therefore the surface brightness and surface density profiles 
should be the same. To check this assumption, we fitted a second set of EFF profiles using only stars fainter 
than the MSTO and brighter than the magnitude for which the completeness is 50\% in the cluster central regions. 
The MSTOs were adopted from Paper~I. The best-fitting EFF profiles are shown in Figures~\ref{fig:effmsto} 
and~\ref{fig:sb419msto} (solid lines). The results are summarized in Table~\ref{tab:sum_eff} (lower section). The profiles 
are much less scattered and smoother toward the outer region than in the first set of profiles. The reason for the 
smoother profiles is probably that the number of stars contributing to the surface brightness is larger, and 
hence statistical fluctuations are smaller. 

Cluster core 2-body relaxation times range from 1-2~Gyr on upwards, and thus mass segregation from this process is only 
expected to be a factor in the densest and oldest clusters in the sample. Only in NGC\,121 differences are 
seen at a possibly significant level in the expected sense of larger core radii for the low mass stars  
exceeding the combined errors by more than a factor of three. Thus we find no compelling evidence for mass 
segregation in the seven intermediate mass clusters in our sample, although we emphasize that deeper observations 
are needed for a definitive test. 

Both for the number density and surface-brightness distributions King profiles were fitted, which are summarized
in the upper and middle sections of Table~\ref{tab:sum_king} for the sample containing all stars above the 
50\% completeness levels and in the lower section for all stars between the 50\% completeness levels and the 
MSTOs.

\begin{figure}
  \epsscale{1.3}
  \plotone{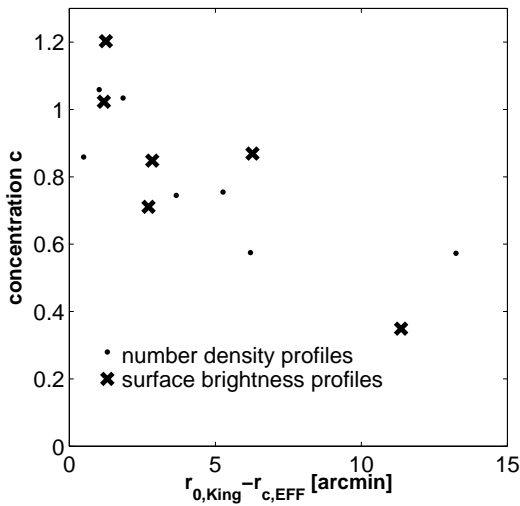}
  \caption{Difference between the King profile scale radius $r_0$ and the core radius $r_c$ from EFF profiles as a function of the concentration parameter c.}
\label{fig:cvsdiff}
\end{figure}

\subsection{Angular distribution and ellipticity}
\label{sec:angluar_dist}

To calculate the projected angular distribution we chose all stars around $C_{phot}$ that lie entirely inside a circle  
on the image (see Table~\ref{tab:sum_ell}). Because the clusters are centered in the upper chip of the camera and the 
tidal radii lie outside the ACS images, this restriction was necessary to avoid artificial fluctuations. Our cluster samples were
subdivided into 12 degree sectors and plotted against the azimuthal angle $\theta$. 
We used a maximum-likelihood approach (McLaughlin et al. 1994, 1995) to obtain a solution for the ellipticity $\epsilon$. 
The observed number density for an intrinsically elliptical distribution of points sampled in {\it circular} annuli is given by

\begin{equation}
\sigma(R,\theta) =  kR^{-\alpha}[cos^2(\theta - \theta_p) + (1 - \epsilon)^{-2}sin^2(\theta - \theta_p)]^{-\alpha/2} + \sigma_b  \\ 
\end{equation}

where $\theta_p$ is the position angle of the major axis, $\sigma_b$ the background density, and the radial falloff is 
modeled as a simple power law with an exponent $\alpha$. 

\begin{figure*}
  \epsscale{1.2}
  \plotone{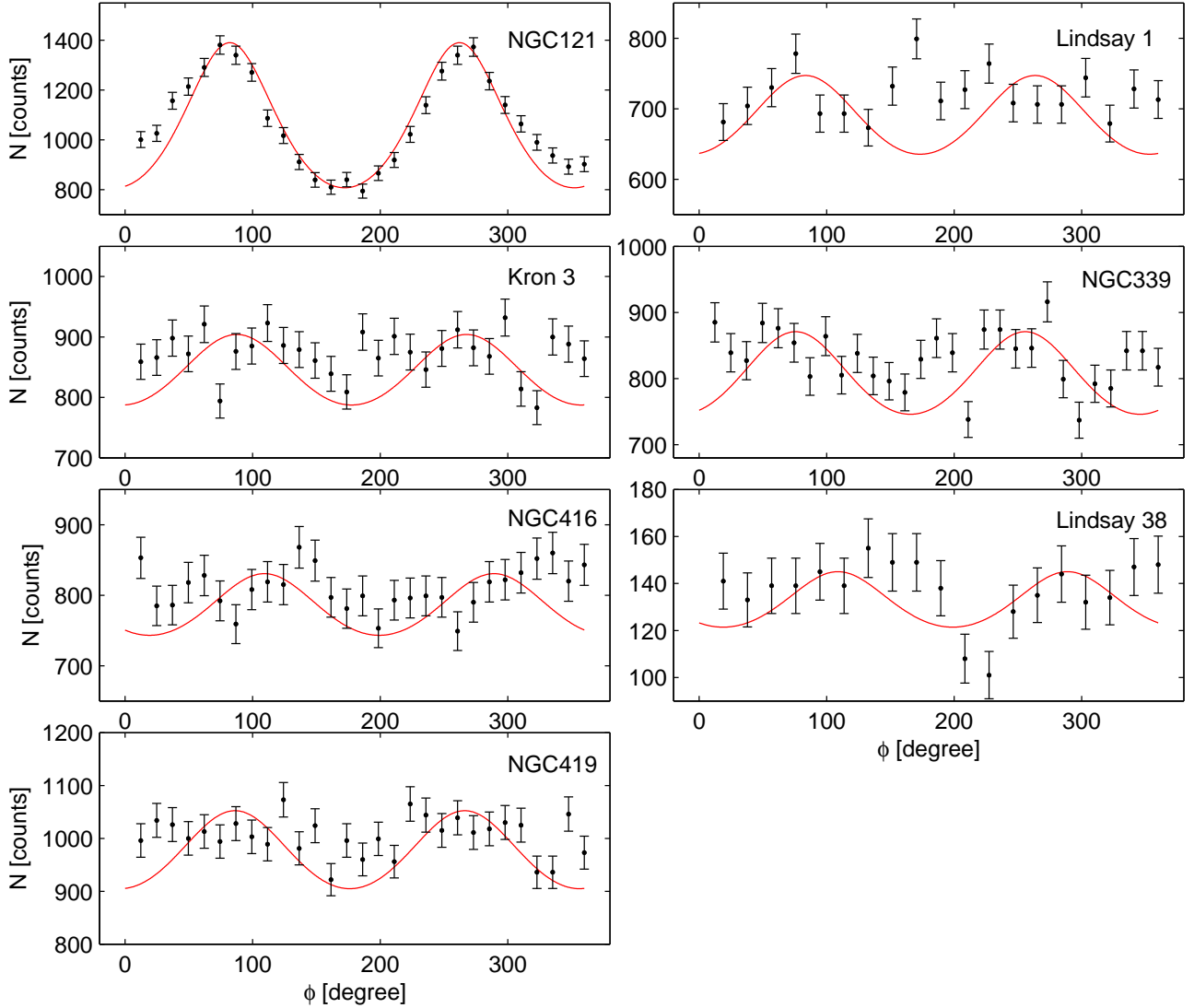}
  \caption{The angular distribution of member stars that lie entirely on the ACS images around the determined
  center of gravity. The solid red curves are the best fits of Eq.(3) of the text, indicating the ellipticity of the clusters.
  The measured ellipticities for all clusters except the one for NGC\,121 should be considered as upper limits, because
  the data could be easily fit with a lower ellipticity and a similarly low $\chi^2$.}
\label{fig:angdist}
\end{figure*}

To get a constraint on $\sigma_b$ we use the distribution of stars in the color-magnitude diagrams (CMDs) of the clusters. 
Because the SMC is not crowded and because it is located at high Galactic latitude \citep[e.g., ][]{ratna85}, we do not 
expect field star contamination levels to be a significant problem. We select those objects directly in the cluster 
CMDs that are definitely not cluster members (carefully excluding e.g. asymptotic giant branch (AGB) and blue straggler star 
(BSS) stars). We use the number of these objects to estimate the background. The normalized background contaminations 
$\sigma_b$ are then used to determine the ellipticities. Equation 3 was fitted to each of our clusters via 
$\chi^2$ minimization. The measured ellipticities for all clusters except NGC\,121 should be considered as upper 
limits, because the data could be easily fit with a lower ellipticity and a similarly low $\chi^2$.
Because we cannot measure the ellipticities at the tidal radii, the values presented here refer to the light
distribution near $r_h$. The angular distributions are shown in Figure~\ref{fig:angdist} and the results are
summarized in Table~\ref{tab:sum_ell}. 

\begin{deluxetable*}{cccccccc}
\tablecolumns{8}
\tablewidth{0pc}
\tablecaption{Ellipticities}
\tablenote{The cluster 3d-distances from the SMC center were calculated in Paper~I. R lists the radii within which 
all stars for the angular distribution calculation were considered. The ellipticities come from this paper and 
the literature values were adopted from \citet{hill06}. All ellipticities determined in this study except the one 
for NGC\,121 should be considered as upper limits. The ages are taken from Paper~I,~II and 
\citet{alves99,crowl01, piatti01,sabbi07}.}
\tablehead{
\colhead{Cluster} & \colhead{Distance from } & \colhead{age} & \colhead{R} & \colhead{$\epsilon$} & \colhead{$\theta_p \pm 180$} & \colhead{$\alpha$} & \colhead{$\sigma_b$}  \\
\colhead{} & \colhead{SMC center [kpc]} & \colhead{Gyr} & \colhead{arcmin} & \colhead{} & \colhead{degree} & \colhead{} & \colhead{$arcmin^2$}} 
\startdata
NGC\,121     & $8.76 \pm 1.1$   & $10.5 \pm 0.5$  & 1.17 & $0.27 \pm 0.06$  & $83$  & $1.77$ & $22.48$\\
Lindsay\,1   & $13.28 \pm 1.0$  & $7.5 \pm 0.5$   & 1.52 & $0.16$           & $83$  & $0.96$ & $23.85$ \\
Kron\,3      & $7.19 \pm 1.1$   & $6.5 \pm 0.5$   & 1.22 & $0.14$           & $88$  & $0.96$ & $28.80$\\
NGC\,339     & $0.73 \pm 2.0$   & $6 \pm 0.5$     & 1.46 & $0.17$           & $76$  & $0.90$ & $62.55$\\
NGC\,416     & $3.94 \pm 1.4$   & $6 \pm 0.8$     & 1.04 & $0.17$           & $109$ & $0.90$ & $264.33$\\
Lindsay\,38  & $6.27 \pm 1.3$   & $6.5 \pm 0.5$   & 1.18 & $0.21$           & $109$ & $0.88$ & $17.82$\\
NGC\,419     & $10.83 \pm 1.6$ & $1.2-1.6$        & 1.14 & $0.14$           & $86$  & $1.10$ & $90.45$\\
&&&&&&& \\
\textit{Literature Sample} &&&&&&& \\
NGC\,411     & $11.1 \pm 1.3$   & $1.2 \pm 0.2$ & - & 0.08 & $ - $ & $ - $ & $ - $ \\
NGC\,152     & $5.58 \pm 1.3$   & $1.4 \pm 0.2$ & - & 0.23 & $ - $ & $ - $ & $ - $ \\
Kron\,28     & $14.78 \pm 1.3$  & $2.1 \pm 0.5$ & - & 0.30 & $ - $ & $ - $ & $ - $ \\
Kron\,44     & $4.37 \pm $ 1.3  & $3.1 \pm 0.8$ & - & 0.26 & $ - $ & $ - $ & $ - $ \\
BS90         & $1.23$	        & $4.3 \pm 0.1$ & - & 0.05 & $ - $ & $ - $ & $ - $ \\
\enddata
\label{tab:sum_ell}
\end{deluxetable*}

\section{Discussion}
\label{sec:discussion}
\subsection{Comparison of the core radii with previous studies}

\begin{deluxetable*}{ccccccccc}
\tablecolumns{8}
\tablewidth{0pc}
\tablecaption{Comparison of $r_c$ [pc] from EFF model fits to literature data}
\tablenote{The core radii of \citet{mackey03b,mclaughlin05,hill06} have been converted to the distances found in Paper~I.}
\tablehead{
\colhead{Reference} & \colhead{NGC\,121}  & \colhead{Lindsay\,1}  & \colhead{Kron\,3} & \colhead{NGC\,339}  & \colhead{NGC\,416} & \colhead{Lindsay\,38} & \colhead{NGC\,419}  \\
\colhead{} & \colhead{} & \colhead{} & \colhead{} & \colhead{} & \colhead{} & \colhead{}  & \colhead{}} 
\startdata
\noindent
This paper           & $3.75 \pm 0.39$ & $13.81 \pm 0.60$ & $6.70 \pm 0.26$ & $8.10 \pm 0.11$ & $2.65 \pm 0.07$ & $9.30 \pm 0.48$ & $2.85 \pm 0.09$   \\    
\citet{carvalho08}   & $2.39 \pm 0.01$ & -                & -               & $7.23 \pm 0.71$  & $2.30 \pm 0.01$ & -                & $2.93 \pm 0.01$ \\
\citet{hill06}       & -               & -                & -               & 8.62             & 3.05            & -                & 4.04            \\
\citet{mclaughlin05} & $3.05 \pm 0.10$ & -                & $6.92 \pm 0.26$ & $7.98 \pm 0.45$  & $3.13 \pm 0.10$ & -                & -               \\
\citet{mackey03b}    & $3.02 \pm 0.10$ & -                & $6.07 \pm 0.18$ & $7.05 \pm 0.30$  & $2.84 \pm 0.10$ & -                & -               \\  
\enddata
\label{tab:comparison_eff}
\end{deluxetable*}

The only study of structural parameters of SMC star clusters based on space-based observations (HST/WFPC2) was 
presented by \citet{mackey03b}. They used imaging data of 10 populous star clusters. Four of these clusters are also 
included in 
the present sample. The most recent ground-based study of structural parameters was published by \citet{carvalho08} 
and is based on data taken with the Danish 1.54~m telescope at the European Southern Observatory, La Silla, Chile. 
These authors studied surface photometry 
of 25 SMC star clusters, of which four overlap with our sample. \citet{hill06} used data from the Magellanic Clouds 
Photometric Survey (MCPS) \citep{zaritsky02}. Structure parameters were derived from fitting both King and EFF 
profiles for 204 star clusters. \citet{mclaughlin05} fitted both King and EFF models to star-count data for 
clusters in the SMC of which four overlap with the present sample. The observed profiles come from combining
the ground-based photographic data from \citet{mackey03b} with data from \citet{kontizas82} and \citet{kontizas83}.
Earlier studies using number-density profiles were published by \citet{kontizas90} 
and are based on photographic plates obtained with the 1.2~m U.K. Schmidt Telescope in Australia. Five clusters of 
the present sample were also studied by \citet{kontizas90}, but owing to the different resolution and depth of the
shallower photographic plates we do not discuss these results here. As noted above, although often called the
'core radius', the King profile scale radius $r_0$ is larger than the true core radius $r_c$, with the difference
being large for lower central concentrations. This effect is illustrated in Fig.~\ref{fig:cvsdiff} where we plot the
difference between $r_0$ and $r_c$ (derived from the EFF profile fits) against central concentration. We must therefore
be careful to compare 'like with like', and we have thus compared our core radii from the EFF profile fits with 
literature data that also used EFF profile fits.

In Table~\ref{tab:comparison_eff} we compare the core radii from EFF profiles from surface-brightness profiles with the
above mentioned previous studies. The EFF core radius of NGC\,121 found in our study is $\sim$1.4~pc larger than the 
corresponding one published by \citet{carvalho08} and is $\sim$0.75~pc larger than the one found by \citet{mackey03b} 
and~\citet{mclaughlin05}. The core radius of NGC\,339 is $\sim$0.9~pc ($\sim$11\%) larger than 
the radius published by \citet{carvalho08} (1 pc = 3.4''), while the core radii of NGC\,416 and Kron\,3 are in good 
agreement. The core radii of Kron\,3 and NGC\,339 are $\sim$0.6~pc ($\sim$10\%) larger than the ones found by 
\citet{mackey03b} while the core radius of NGC\,416 agrees well with our value. The core radii of Kron\,3 and NGC\,339 
are in very good agreement with \citet{mclaughlin05}, but the core radius of NGC\,416 is $\sim$0.5~pc ($\sim$18\%) smaller. 
Comparing our result with \citet{hill06} we find that their core radii for NGC\,339, NGC\,416, and NGC\,419 are all larger
than the values of this study by about 0.5~pc, 0.4~pc, and 1.5~pc, respectively. 

Comparing the half-light radii for NGC\,121, Kron\,3, and NGC\,339 determined by estimating $L_{tot}$ using surface-brightness 
data to those radii published by \citet{mclaughlin05}, we find that their half-light radii of NGC\,121, Kron\,3, and NGC\,339 
are $\sim$16\% (5.96~pc), $\sim$27\% (10.54~pc), and $\sim$26\% (12.65~pc) smaller than those found in this study. 
The remaining clusters of the present sample do either not overlap or the half-light radii could not be calculated. 

In the LMC, \citet{mackey03a} found evidence for double or post-core-collapse (PCC) clusters in the surface brightness 
profiles among the oldest clusters, as well as in one globular cluster in the Fornax dwarf spheroidal galaxy 
\citep{mackey03c}. A PCC cluster is characterized by an apparent power-law profile in its innermost region, which is
different from a constant-density core as found in the EFF and King profiles. No evidence of this kind 
of clusters was found for SMC clusters \citep{mackey03b} or in our study. 

\begin{figure}
  \epsscale{1.2}
  \plotone{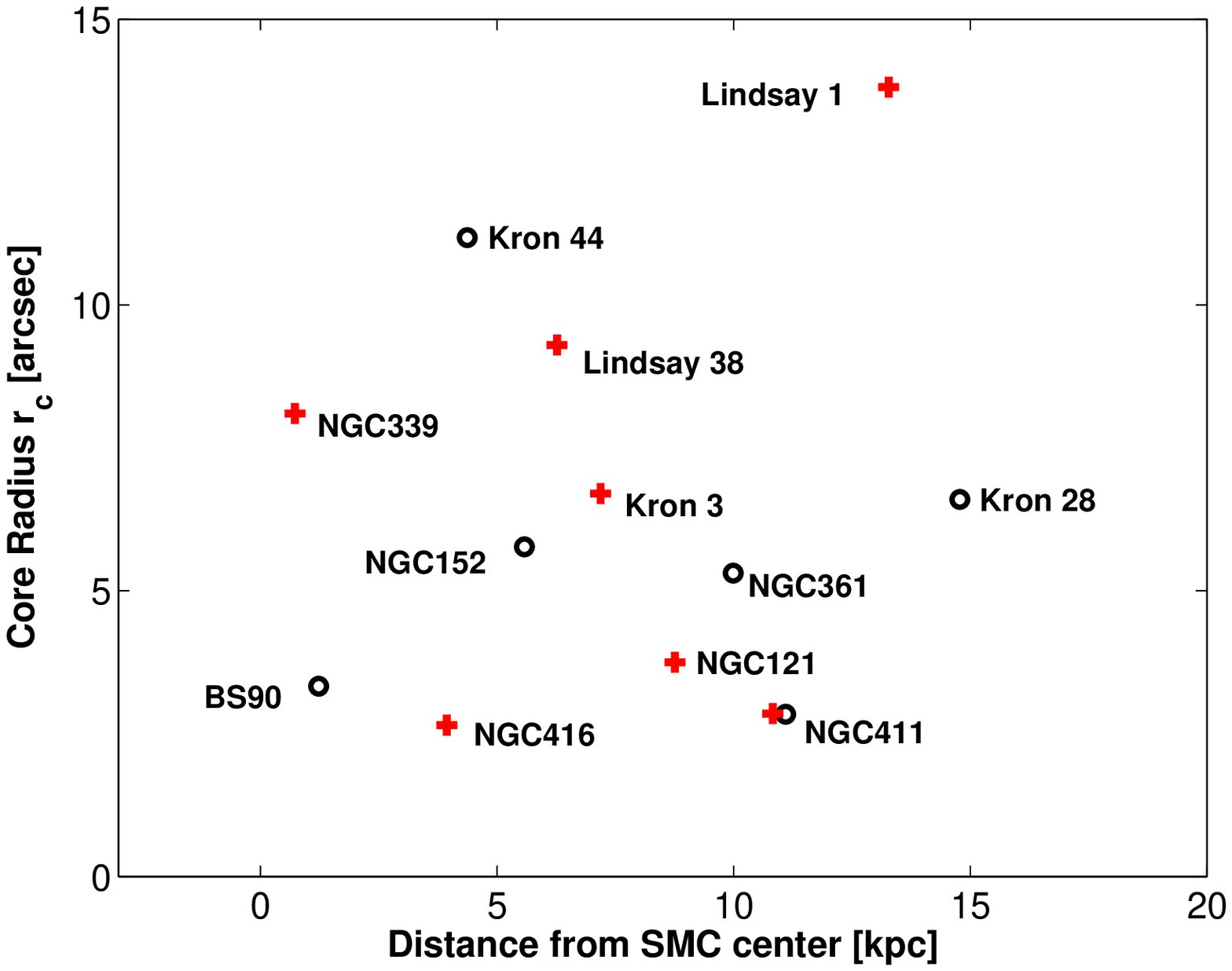}
  \caption{Core radii vs distance from the SMC center for the clusters in our sample (plus) and clusters for which 
  reliable distances were found in the literature (circles). The distances were adopted from Paper~I and \citet{crowl01}. 
  The core radii for NGC\,152, NGC\,361, and NGC\,411 are taken from \citet{mackey03b}, while for Kron\,28 and Kron\,44 
  the values are taken from \citet{hill06}.}
\label{fig:rcvsdist}
\end{figure}

One additional uncertainty in the inter-comparisons of physical cluster core and tidal radii is the
distance modulus. 
In previous studies a single distance modulus was assumed and applied to the calculations. However, the SMC may 
have a depth extent of up to 20~kpc \citep{Math88,Hatz93,crowl01,lah05}. With the exception of NGC\,419, for which 
we used isochrone fitting, we determined the distance modulus for each cluster using the red clump magnitude 
\citep{glatt08a}. However, no correlation between core radius and distance from the SMC center was found 
(Fig.~\ref{fig:rcvsdist}). 

\subsection{Age-radius relation}

\begin{figure}
  \epsscale{1.2}
  \plotone{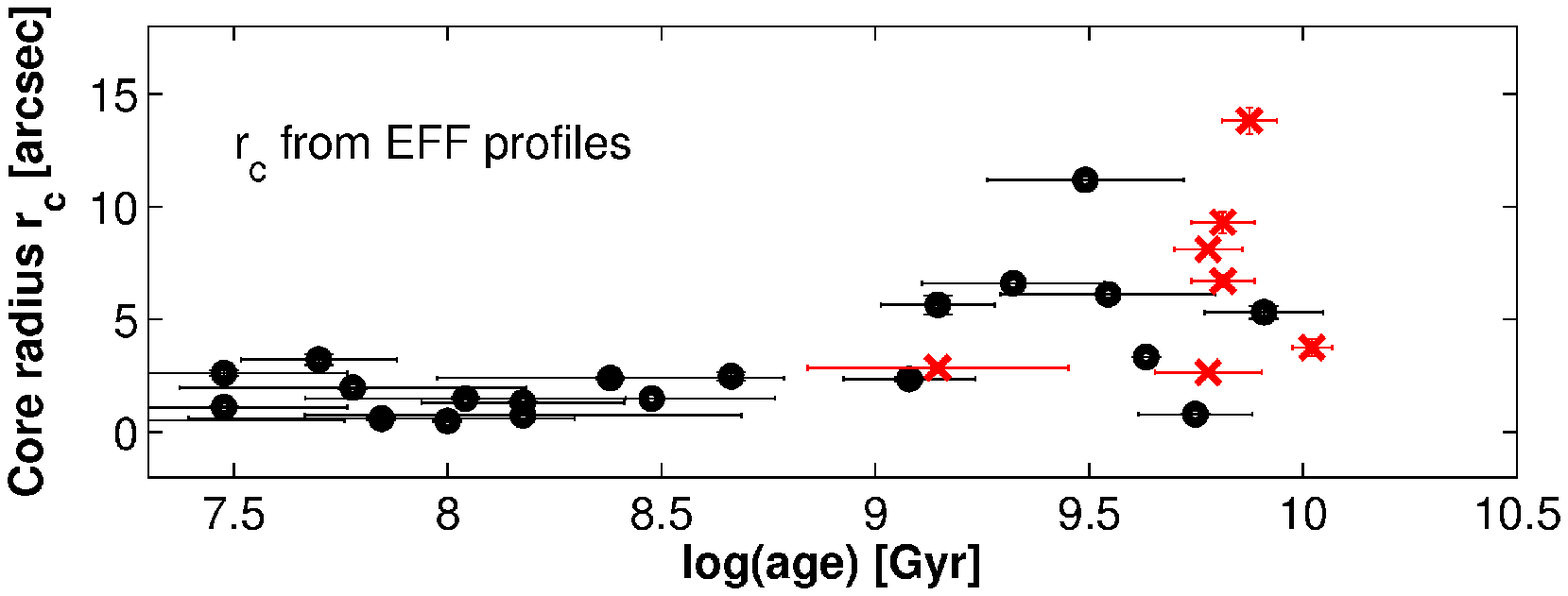}
  \caption{Age vs core radius $r_c$ of our cluster sample (crosses) and additional sample found in the literature 
  (circles). The adopted core radii and ages are listed in Table~\ref{tab:lit_radius}. Because the errors of the
  EFF core radii in both our and in the literature sample are very small, the errorbars are barely visible.}
\label{fig:agevsrc}
\end{figure}

In the LMC, a trend for larger core radii with increasing age has been found \citep{mackey03a} and evidence for the same
trend appears to be present in the SMC \citep{mackey03b}. In their studies, core radii of 10 SMC star clusters and 53 
LMC clusters were
determined using EFF profiles and compared to each other. The SMC clusters may have slightly larger core
radii on average than the LMC clusters, but the authors claimed that this could be the result of uncertainties in 
both the SMC distance modulus (for all clusters a distance modulus of 18.9 was used) and the large depth extension
of the SMC. The youngest SMC and LMC clusters all had compact cores, whereas older clusters showed a bifurcation with 
most clusters following a lower sequence and some clusters exhibiting increased core radii. 

Combining our results and literature values (Tab.~\ref{tab:lit_radius}), we confirm the proposed relationship between 
cluster age and spread in core size in the SMC. In Figure~\ref{fig:agevsrc} core radii from EFF 
profiles are shown. The trend of older clusters having a larger range in core 
radii than the younger population is clearly visible.  The oldest star cluster, NGC\,121, has a small
core radius of $3.75 \pm 0.39$~pc, while for the second oldest cluster, Lindsay\,1, the core radius is rather large 
with $13.81 \pm 0.60$~pc. One of our intermediate-age clusters has a radius larger than 10~pc (Lindsay\,1). The intermediate-age 
cluster Kron\,44 has the largest core radius of the literature sample 
($r_c$ = 11.18~pc; Hill et al. 2006). Its age is $3.1 \pm 0.8$~Gyr \citep{piatti01}. Star clusters younger than 1~Gyr 
have core radii smaller than 4~pc. Nevertheless, only Lindsay\,1 and Kron\,44 have significantly 
larger core radii than the rest of the clusters considered here. The clusters NGC\,361, NGC\,152, Kron\,28, Lindsay\,11, 
Kron\,3, NGC\,339, and Lindsay\,38 have radii between 5 and 8.5~pc, which is slightly larger on average than the core radii of the 
remaining 21 clusters, whose core radii are smaller than 5~pc. Figure~\ref{fig:agevsrc} suggests that among the older 
clusters some objects seem to have experienced a significant change in $r_c$, while for others $r_c$ remained almost 
unaltered.

For GCs, the concentration parameter c traditionally is around or even larger than 1, implying a compact isothermal 
central 
region and an extended tidally truncated outer region, while open clusters have $c < 1$, which is an indication that open 
cluster are more diffuse objects \citep[e.g., ][]{harris96,binney98,bonatto08}. The oldest and only globular cluster
in the SMC, NGC\,121, has a concentration parameter of $c = 1.034 \pm 0.12$. For comparison, the Galactic globular clusters 
47~Tuc, 
NGC\,288, and NGC\,6909, all non-PCCs, have c = 2.04, c = 0.98, and c = 0.76 \citep{harris96}. The youngest cluster 
in the present sample, NGC\,419, with an age of 1.2-1.6~Gyr, has a globular cluster-like concentration parameter of 
$c = 1.059 \pm 0.08$. This is the case for all three profiles. In its CMD we see indications of a double
or even multiple main sequence turnoffs (see Paper~I), a feature seen also in the two LMC star clusters, NGC\,1846 and 
NGC\,1806, known to have a double main-sequence turn-off (Mackey et al. 2008, see also Mackey \& Broby Nielson 2007). 

Plotting mass versus core radius does not show any correlation for LMC and SMC star clusters \citep{mackey03a,mackey03b},
and no significant difference between masses of young and old clusters was found. If only massive star clusters had 
large core radii one could argue that the younger low-mass clusters dispersed after a few Gyr, but at least 
in the LMC and SMC this does not seem to be the case. Since the age-radius correlation has been observed in the combined
sample of SMC, LMC, Fornax, and Sagittarius star clusters, \citet{mackey03c} emphasized the possibility of a universal 
physical process as the origin of this trend. While our results are consistent with this possibility, additional 
observations and theoretical studies are needed to establish if an internal process is at work. 

\subsection{Cluster evolution}

\begin{deluxetable}{ccccc}
\tablecolumns{5}
\tablewidth{0pc}
\tablecaption{Literature Sample}
\tablenote{The cluster ages we adopted from: (1) \citet{hodge83}, (2) \citet{hodge87}, (3) \citet{elson85}, 
(4) \citet{ahumada02}, (5) \citet{daco98}, (6) \citet{sabbi07}, (7) \citet{chiosi06}, (8) \citet{piatti05}, 
(9) \citet{mighell98}, (10) \citet{alves99}, (11) \citet{crowl01}, (12) \citet{piatti01}, (13) \citet{mould92}, 
and the EFF core radii from: (14) \citet{mackey03b} and (15) \citet{carvalho08}.}
\tablehead{
\colhead{Cluster} & \colhead{age} & \colhead{$r_{c,EFF}$} & \colhead{References} \\
\colhead{} & \colhead{Gyr} & \colhead{pc} & \colhead{}  }
\startdata
NGC\,176        & $0.46 \pm 0.01$  & $2.48 \pm 0.19$ & 1, 14  \\
Kron\,17        & $0.30 \pm 0.10$  & $1.49 \pm 0.01$ & 2, 15  \\
$NGC\,241+242$  & $0.07 \pm 0.04$  & $0.60 \pm 0.07$ & 3, 15  \\
NGC\,290        & $0.03 \pm 0.01$  & $1.11 \pm 0.03$ & 4, 15  \\
Lindsay\,48     & $0.15 \pm 0.04$  & $1.32 \pm 0.02$ & 1, 15  \\
Kron\,34        & $0.24 \pm 0.12$  & $2.39 \pm 0.04$ & 3, 15  \\
NGC\,330        & $0.03 \pm 0.01$  & $2.61 \pm 0.12$ & 5, 14  \\
Lindsay\,56     & $0.006 \pm 0.01$ & $0.51 \pm 0.01$ & 4, 15  \\
NGC\,346        & $\sim 0.003$     & $2.01 \pm 0.03$ & 6, 6  \\
IC\,1611        & $0.11 \pm 0.05$  & $1.49 \pm 0.06$ & 3, 15  \\
IC\,1612        & $\sim 0.10$	   & $0.49 \pm 0.01$ & 7, 15  \\
Lindsay\,66     & $0.15 \pm 0.10$  & $0.74 \pm 0.01$ & 8, 15  \\
NGC\,361        & $8.10 \pm 1.20$  & $5.31 \pm 0.28$ & 9, 14  \\
Kron\,47        & $\sim 0.007$     & $1.74 \pm 0.15$ & 7, 15  \\
IC\,1624        & $0.06 \pm 0.03$  & $1.97 \pm 0.03$ & 3, 15  \\
NGC\,411        & $1.20 \pm 0.20$  & $2.84 \pm 0.11$ & 10, 14 \\
NGC\,458        & $0.05 \pm 0.01$  & $3.22 \pm 0.24$ & 1, 14  \\
Lindsay\,114    & $5.60 \pm 0.50$  & $0.80 \pm 0.02$ & 4, 4  \\
NGC\,152        & $1.4 \pm 0.20$   & $5.77 \pm 0.42$ & 11, 14 \\
Kron\,28        & $2.1 \pm 0.50$   & $6.60$	     & 12, 15 \\
Kron\,44        & $3.1 \pm 0.80$   & $11.18$	     & 12, 15 \\
Lindsay\,11     & $3.5 \pm 1$	   & $6.11$	     & 13, 15 \\
BS90            & $4.3 \pm 0.10$   & $3.33$	     & 6, 15  \\
\enddata 
\label{tab:lit_radius}
\end{deluxetable}

It is intriguing that LMC and SMC clusters seem to have experienced a similar structural evolution, even though the 
two galaxies
show strong differences in various other aspects. The SMC contains only one old GC, NGC\,121, which is 2-3~Gyr younger 
than the oldest GC in the LMC and MW (Paper~II). The second oldest SMC star cluster, Lindsay\,1, has an age of 
$7.5 \pm 0.5$~Gyr, and since then compact populous star clusters have formed fairly continuously until the present day 
\citep[e.g.,][]{daco02,glatt08a}. Furthermore, the intermediate-age clusters in the SMC appear to be capable of surviving
for a Hubble time, due to their high mass and the structure of the SMC (no bulge or disk to be passed; Hunter et al. 2003;
Lamers et al. 2005; Gieles et al. 2007). However, the SMC has a moderately dense ''bar'' and we do not know how its 
clusters orbit in the SMC. 

In contrast 
to the SMC, the LMC had two main epochs of cluster formation \citep[e.g., ][]{Bertelli92} and a well-known ''age-gap'' 
between $\sim$4-9~Gyr \citep[e.g.,][]{holtz99, john99, Harris01}, in which no (surviving) star clusters have formed.  
Several GCs are found with coeval ages like the Galactic GCs and GCs in other dwarf galaxies \citep[e.g., ][]{ols91,
olsen98,john99,grebel04}. We know only of one LMC star cluster, ESO 121-SC03, that has an age of 8.3-9.8~Gyr 
\citep{mackey06}, which defines the lower limit of the old LMC star cluster age distribution. 

In the SMC, \citet{hill06} found the distribution of cluster core sizes to be broader than in the MW, which they argue 
is due to a prevalence of surviving low-concentration clusters in the SMC. In her analysis of the LMC star clusters, 
\citet{elson91} noted that for clusters of a given age, there appears to exist an upper limit for their core size. 
Moreover, this limit was found to increase with age, which was later confirmed by the analysis of \citet{mackey03a}. 
Young clusters are observed to have very compact cores of e.g. $\sim$ 2.13~pc (NGC\,1711) and $\sim$ 1.33~pc (NGC\,1805), 
whereas the cores of older clusters can reach extents of up to 13~pc. 

N-body 
simulations of \citet{goodwin06} illustrate how the structural parameters of star clusters change with time. A major driver 
of these changes is the expulsion of gas, which was not converted into stars via star formation. The minimum local star 
formation efficiency to leave a bound massive star cluster is $\sim$25\%, and higher efficiencies are possible 
\citep[e.g., ][]{parmentier08a}. But even in efficient cases, a significant amount of unused gas remains. 
\citet{kroupa02} suggested that populous star clusters expel their unused gas explosively due to the presence of numerous 
O stars during their early evolutionary stages. \citet{goodwin06} also suggested a rapid gas removal caused by stellar 
winds and supernovae during the first 20~Myr. As a consequence the young clusters may find themselves out of virial
equilibrium; the stars have too large a velocity dispersion for the new, reduced gravitational potential. In order 
to re-stabilize, the cluster expands on a few crossing times scales of $<$10~Myr (see e.g., Sabbi et al. 2007).
The most extreme core radii can be explained by external processes (e.g., for clusters that are not isolated), or by 
radically different stellar populations (mergers of clusters and effects from variable tidal fields, Mackey \& Gilmore 
2003a). 

Various estimates of characteristic cluster disruption time-scales in different star cluster environments 
have been calculated \citep[e.g., ][]{gieles08,parmentier08b}. Many studies were based on a constant cluster formation
rate (CFR) as a function of time. The poorly understood time-variable CFR of the LMC and SMC complicates such an analysis.
\citet{parmentier08b} analyzed the cluster disruption time-scale in the LMC using Monte Carlo simulations. For younger
clusters (age $<$ 5~Gyr), the general behavior of the CFR is recovered. It has been increasing steadily from about 
0.3 clusters~Myr$^{-1}$ 5~Gyr ago to a present rate of $\sim$25 clusters~Myr$^{-1}$. For older clusters (age $>$ 
5~Gyr), the CFR is very uncertain. It is possible that the CFR has increased steadily over a Hubble-time from $\sim$ 1 
cluster~Gyr$^{-1}$ 13~Gyr ago to its present value. For the SMC such studies have not been published, but the lack of 
very old SMC star clusters shows that the CFR of the LMC and SMC either varies significantly or that the CFR was rather 
constant but most clusters dissolved before reaching intermediate ages. 

\begin{figure}
  \epsscale{1.2}
  \plotone{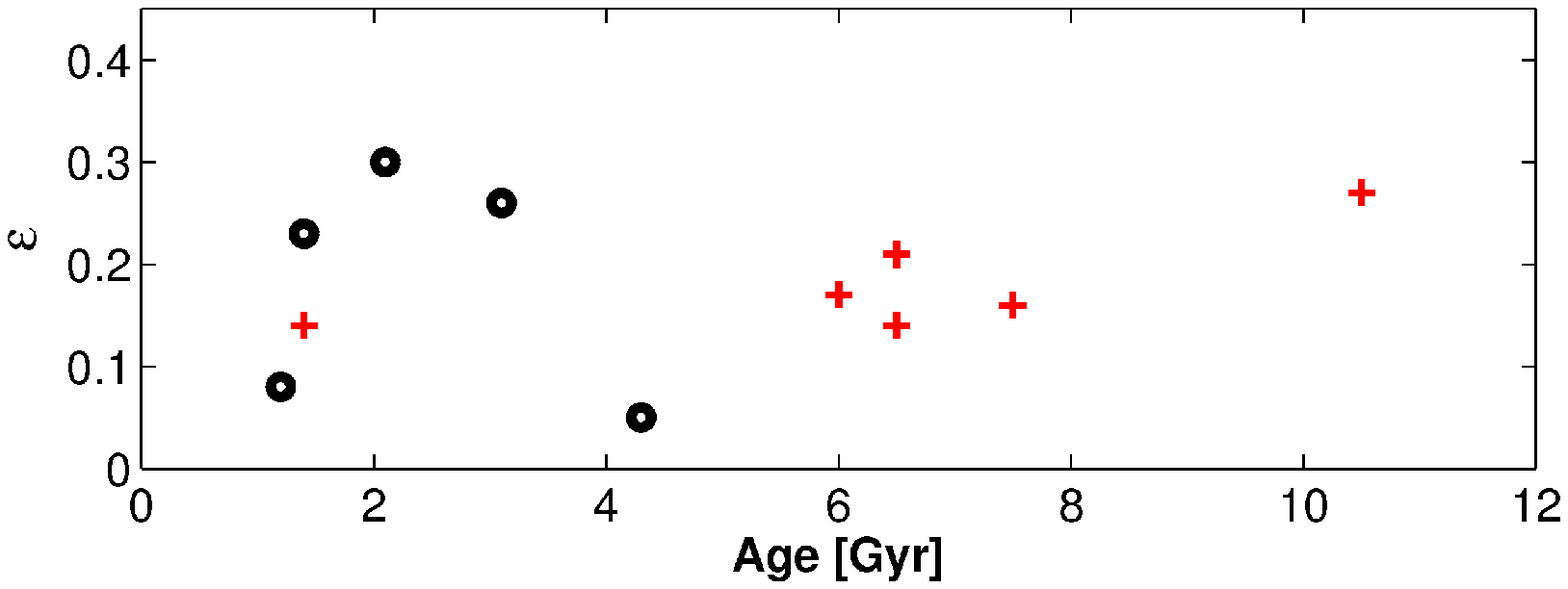}
  \caption{Age vs ellipticity showing the clusters from our sample (crosses) and five 
  clusters from the literature (circles). The ellipticities for NGC\,152, NGC\,411, Kron\,28, Kron\,44, and BS90 are 
  taken from \citet{hill06} and the ages from \citet{alves99,crowl01,piatti01,sabbi07}. It seems that the younger 
  clusters tend to be flatter than the older clusters, although NGC\,121 is the oldest and the flattest cluster. The
  measured ellipticities in this study, except the one of NGC\,121, should be considered as upper limits.}
\label{fig:ellit}
\end{figure}

\subsection{Ellipticities}
\subsubsection{Comparison with previous studies}

None of the clusters in our sample exhibits a significant flattening except NGC\,121 and Lindsay\,38, the sparsest 
cluster in the present sample.  
Because the tidal radii of our clusters lie outside the ACS images, we have to assume that the ellipticities in 
the clusters' interiors are the same as at the tidal radii. For Lindsay\,38, no ellipticity determination 
was found in the literature. 

Measuring ellipticities is strongly dependent on the correct determination of the cluster center. Background 
and foreground determination, as well as stochastic effects influence the determination of isophotes. However, as
much of the SMC is not crowded and at high Galactic latitude \citep[e.g., ][]{ratna85}, the effects of field star 
contamination should not be severe. Another explanation for the large SMC cluster ellipticities
might be the influence of the local SMC field star background, the different measuring methods as well as the 
differing radii at which the ellipticities were measured. Some Galactic globular clusters have been found
showing increasing ellipticities at larger distances from the cluster core \citep{white87}. 
These might be the reasons for the differences 
between the values presented here and those found by \citet{kontizas90} and~\citet{hill06}.
We have to emphasize that the determination of the ellipticity is quite uncertain,
especially when it is fairly small as it is for most of the present clusters. The results
of the the present study and the literature data are in good agreement (Table~\ref{tab:ellipticity}).

\begin{deluxetable*}{cccccccc}
\tablecolumns{7}
\tablewidth{0pc}
\tablecaption{Comparison of the ellipticities}
\tablenote{The ellipticities by \citet{kontizas90} were measured at the inner-most parts. All ellipticities determined in
this study except the one for NGC\,121 should be considered as upper limits.}
\tablehead{
\colhead{Reference} & \colhead{NGC\,121} & \colhead{Lindsay\,1} & \colhead{Kron\,3} & \colhead{NGC\,339} & \colhead{NGC\,416} & \colhead{Lindsay\,38}  & \colhead{NGC\,419} \\
\colhead{} & \colhead{} & \colhead{} & \colhead{} & \colhead{} & \colhead{} & \colhead{} & \colhead{}} 
\startdata
This paper         & $0.27 \pm 0.06$ & 0.16 & 0.14 & 0.17 & 0.17 & 0.21 & 0.14 \\         
\citet{hill06}     &  -              &  -   &  -   & 0.17 & 0.11 &  -   & 0.09   \\
\citet{kontizas90} & 0.28            & 0.10 & 0.10 & 0.23 & 0.13 &  -   & 0.23 \\						   
\citet{geyer83}    & 0.30 & - & - & - & - & - & - \\
\enddata
\label{tab:ellipticity}
\end{deluxetable*}

\subsubsection{Astrophysical Implications}

The flattening distributions of star clusters in the SMC, LMC and the MW are known to be very different 
\citep[e.g., ][]{kontizas89,han94,goodwin97}. SMC star clusters in their outer regions are typically much more 
flattened than those of the MW and even flatter than those in the LMC \citep{kontizas90}. In Figure~\ref{fig:ellit} 
we show the relation of ellipticity versus age. No evident correlation can be seen. Unfortunately, not enough reliable 
distances have been measured to make an accurate statement about a possible dependence on galactocentric
radius in the LMC and the SMC for a large sample of clusters. We extend our sample by adding five populous star clusters 
for which reliable distances, ages, and ellipticities have been determined elsewhere in the literature 
(see Table~\ref{tab:sum_ell} for references). The younger clusters (age $\lesssim$5~Gyr) seem to be flatter 
(larger $\epsilon$) than the older objects. In the MW, the flatter GCs are located close to the Galactic center, while 
for SMC such a correlation is not visible. Clusters lying behind the SMC center seem to be flatter than 
the ones lying in front. But we have to emphasize that the shown sample is very small and highly incomplete. 

Like the SMC, the LMC does not show a relation between age and ellipticity, while clusters of all ages 
are significantly more elliptical than the Galactic GCs \citep{goodwin97}. The Galactic GCs appear to modify 
their original structure and become less flattened at higher age. \citet{han94} argue that the 
difference between GC ellipticities in the LMC and the Milky Way are caused by the morphologies of the 
parent galaxies. They further showed that the GCs also vary in their shapes: LMC and SMC clusters are well-represented
by triaxial spheroids, while Galactic GCs are oblate spheroids. Young LMC clusters appear to be highly 
flattened. Did the original structure of the older 
Galactic population get modified during their lifetime, and if yes, why did the old LMC and SMC cluster population 
remained unchanged? This may be explained with the different dynamical influence and therefore the varying strength of the 
tidal field of the parent galaxy \citep{vandenbergh08}. A strong tidal field will make the Galactic GCs more
spherical during their orbits around the galaxy. The LMC and SMC have a totally different structure and no bulge 
or disk has to be passed \citep{hunter03,lamers05,gieles07}. The tidal field of the LMC and the SMC might not be 
strong enough to modify the shape of their clusters significantly, which might be the reason for their flat shapes. 
This point  merits further exploration, e.g. via massive star cluster shape studies in nearby starburst and normal
galaxies.

\section{Summary}

We derive structural parameters for the seven SMC star clusters NGC\,121, Lindsay\,1, Kron\,3, 
NGC\,339, NGC\,416, Lindsay\,38, and NGC\,419 based on stellar number density and surface brightness profiles
and HST/ACS stellar photometry. We used King and EFF models to determine core radii, half-light radii, tidal radii, 
concentration parameters, and ellipticities of the star clusters. Half-light radii could only be estimated because 
the tidal radii lie outside the field of view of the ACS images. 
 
Although our sample of SMC clusters is highly incomplete even after adding literature values, we confirm the result 
of \citet{mackey03b} who found an increased scatter in core radii for older clusters (age $>$ 1~Gyr) in dwarf galaxies. 
In the LMC this trend is more apparent \citep{mackey03a}, perhaps a result of the LMC containing a much larger number of 
star clusters than the SMC. 

We find intermediate age star clusters in the SMC to have larger half light radii and smaller concentration 
parameters than typical Galactic globular clusters of similar mass \citep[e.g., ][]{djorgovski94}.  
Indeed some of the clusters in this study could be classified as ``faint fuzzies" based on their sizes and luminosities 
\citep{Sharina05}. Thus these SMC clusters add to the trend for low density galaxies to contain 
older survivor star clusters with relatively high masses with a wide range of central densities extending from 
the dense globular cluster regime to quite low values.  

The cluster formation history of the LMC also appears to be quite different from that of the SMC. The similar cluster 
structural patterns in the two galaxies is therefore intriguing. In the MW, many of the oldest clusters experienced 
modifications of their original structure during their lifetimes, and many of the oldest halo clusters have developed 
cores, probably due to internal processes. One possible additional reason for the differences in the evolution of cluster 
structures between MW and SMC might be the morphology of the host galaxies. Low central concentration clusters can more 
easily survive in the SMC and LMC, 
while in the MW clusters have to pass the Galactic disk or bulge while orbiting the Galaxy. The lack of correlation 
between core radius and distance from the SMC center and the low density of the SMC both suggests that the cluster
structures are little disturbed by external effects within these galaxies and primarly are driven by internal dynamical 
evolution \citep[e.g., ][]{mclaughlin05}.

Our data also show that the inner regions of intermediate-age clusters in this sample have rather spherical shapes, 
while in their outer zones and the oldest
cluster, NGC\,121, have higher ellipticities. Previous studies found higher ellipticities for the intermediate-age 
clusters at larger radii. We find no correlation between outer ellipticity and age, or outer ellipticity and distance 
from the SMC center, where we can take advantage of the 3-D information on the SMC cluster distribution from Paper~I. 
Consistent with our conclusion regarding structures, the shapes of SMC clusters could remain elliptical if they experience 
little externally driven dynamical modification during their lifetimes. 

This study indicates that the properties of the rich SMC star clusters are largely determined by internal processes. 
These objects thus can provide powerful tests for models of the intrinsic dynamical evolution of star clusters, while also serving 
as evolutionary markers for the SMC. Combinations of data from the MW and its satellites will continue to illuminate
the actions of the internal and external astrophysical processes that shape star clusters.

\acknowledgments
We would like to thank an anonymous referee for his or her useful comments. 
We gratefully acknowledge support by the Swiss National Science Foundation through grant 
number 200020-105260 and 200020-113697. Support for the US component of this program GO-10396 
was provided by NASA through a grant from the Space Telescope Science Institute, which is operated 
by the Association of Universities for Research in Astronomy, Inc., under NASA contract 
NAS 5-26555. Gisella Clementini and Monica Tosi have been partially supported by PRIN-MIUR-2004 
and PRIN-INAF-2005. Andreas Koch acknowledges support by an STFC postdoctoral fellowship and 
Jay Gallagher also obtained helpful additional support from the 
University of Wisconsin Graduate School and from the Heidelberg Graduate School of
Fundamental Physics within the framework of the Excellence Initiative by the German Research
Foundation (DFG) through grant number GSC 129/1.

\clearpage

\end{document}